\documentclass[
    aps,
    prb,
    reprint,
    superscriptaddress,
    amsmath,
    amssymb
]{revtex4-2}

\usepackage{graphicx,color}
\usepackage{amsmath}
\usepackage{amssymb}
\usepackage{natbib}
\usepackage{array}
\usepackage{tabularx}
\usepackage{multirow}
\usepackage[colorlinks=true,linkcolor=blue,citecolor=blue,urlcolor=blue]{hyperref}
\usepackage{placeins}
\usepackage{comment}
\usepackage{cellspace}
\usepackage{float}

\usepackage{newtxtext}
\usepackage{newtxmath}
\usepackage{bm}

\newcommand{\dto}{Dy$_2$Ti$_2$O$_7$}
\newcommand{\hto}{Ho$_2$Ti$_2$O$_7$}
\usepackage{xcolor}

\usepackage[normalem]{ulem}

\begin{document}

\title{Demagnetization effect on magnetic noise measurements in spin ice materials}

\author{F. Morineau} 
\email[]{felix.morineau@neel.cnrs.fr}
\affiliation{Institut N\'eel, CNRS, Universit\'e Grenoble Alpes, 38042 Grenoble, France}
\author{C. Paulsen} 
\affiliation{Institut N\'eel, CNRS, Universit\'e Grenoble Alpes, 38042 Grenoble, France}
\author{G. Balakrishnan} \affiliation{Department of Physics, University of Warwick, Coventry, CV4 7AL, United Kingdom}
\author{D. Prabhakaran} \affiliation{Department of Physics, Clarendon Laboratory,  Oxford University, Oxford, OX1 3PU, United Kingdom}
\author{K. Matsuhira}\affiliation{Kyushu Institute of Technology, Kitakyushu 804-8550, Japan}
\author{S. R. Giblin}
 \affiliation{School of Physics and Astronomy, Cardiff University, Cardiff CF24 3AA, UK}
\author{E. Lhotel}
\email[]{elsa.lhotel@neel.cnrs.fr}
\affiliation{Institut N\'eel, CNRS, Universit\'e Grenoble Alpes, 38042 Grenoble, France}

\begin{abstract}
Magnetic noise spectroscopy provides direct access to spontaneous time-dependent magnetization fluctuations in correlated magnetic systems, including spin liquids, spin ices, and spin glasses. Here we investigate how demagnetizing fields shape the magnetic noise spectra measured in spin ice. By combining magnetic noise and ac susceptibility measurements on single-crystal spin ice samples with different sizes and shapes, we show that magnetic noise, like linear response, is strongly renormalized by sample geometry. As a consequence, measured noise spectra do not directly yield the intrinsic fluctuation spectrum, and demagnetization corrections cannot in general be determined from noise data alone. Instead, sample geometry must be treated as a central experimental control parameter for accessing intrinsic dynamics. These results establish the role of boundary conditions in fluctuation spectroscopy and provide a framework for quantitatively comparing noise measurements with microscopic theories of spin dynamics.
 \end{abstract}

\maketitle

\section{Introduction}
Understanding the dynamical properties of magnetic excitations is central to identifying and characterizing exotic states of matter, including frustrated magnets and spin liquids. In such systems, strong correlations and local constraints suppress conventional long-range order, shifting the focus from static order parameters to dynamical correlations. The fundamental quantity of interest is the time-dependent magnetization autocorrelation function, which encodes the nature of the low-energy excitations and their collective dynamics. Experimental access to these dynamics is traditionally obtained through probes such as neutron scattering, time-dependent magnetic relaxation, and ac susceptibility measurements. 
While powerful, many of these techniques infer the system's dynamics through its response to an applied perturbation or through response-related correlation functions, rather than by directly detecting its spontaneous fluctuations. As a result, they provide indirect access to dynamical correlations, rather than a direct measurement of the intrinsic fluctuation spectrum itself.

Magnetic noise spectroscopy offers a complementary and, in some respects, more fundamental approach. By directly detecting spontaneous magnetization fluctuations using ultrasensitive SQUID sensors, it provides access to the magnetization autocorrelation function without externally driving the system. Originally developed in the context of spin glasses~\cite{Ocio85, Herisson02}, the technique has recently experienced a revival driven by substantial advances in SQUID sensitivity and bandwidth, enabling the detection of extremely weak magnetic signals over a broad frequency range. In systems lacking conventional order parameters including spin liquids such as spin ice~\cite{Dusad19, Samarakoon22, Morineau25, Takahashi25} this ability to measure intrinsic fluctuations is particularly valuable, as it enables direct comparison between experimentally measured fluctuation spectra and microscopic dynamical models~\cite{Hallen22}.

When magnetic noise measurements are combined with ac susceptibility, one can test the fluctuation-dissipation relation (FDR), which links spontaneous fluctuations to linear response in thermal equilibrium~\cite{Kubo66}. The FDR provides a stringent and model-independent benchmark: agreement confirms equilibrium dynamics and internal consistency, while deviations signal genuine nonequilibrium behavior. In a recent work~\cite{Morineau25}, we applied this framework to the low-temperature regime of spin ice, where nonequilibrium phenomena have been reported~\cite{Snyder04, Paulsen14}. We found that the FDR is satisfied in the equilibrium regime only when the measured spectral density $S_{\rm meas}$ is compared with the measured susceptibility $\chi_{\rm meas}$, rather than with the intrinsic susceptibility. This observation implies that magnetic noise, like susceptibility, is influenced by demagnetizing fields and is therefore not a purely intrinsic quantity.

It is well established in the spin ice literature that ac susceptibility must be corrected for demagnetizing effects in order to extract intrinsic material properties~\cite{Matsuhira01, Quilliam11, Bovo13, Twengstrom17}. These effects arise from the magnetic field generated by the sample's own magnetization and depend strongly on sample geometry through the demagnetizing factor $N$. That the FDR holds only at the level of measured quantities indicates that magnetic noise is likewise shaped by these boundary conditions: $S_{\rm meas}$ is a geometry-dependent observable. This conclusion is initially counterintuitive; magnetic noise is measured in nominally zero applied field, where no macroscopic magnetization would seem to require correction. However, the relevant quantity is not the mean magnetization but its fluctuations: spontaneous magnetization fluctuations generate internal fields whose spatial distribution is constrained by the sample geometry. A recent theoretical treatment by Bramwell~\cite{Bramwell25} has clarified the microscopic origin of these effects and established quantitative predictions for how demagnetizing fields renormalize fluctuation spectra.

Here we provide an experimental test of these predictions by systematically investigating the influence of the demagnetizing factor $N$ on magnetic noise measurements. Using ac susceptibility as a calibrated reference and the FDR as a stringent internal consistency condition, we measure both response and fluctuations in the same experimental configuration, enabling a direct, parameter-free comparison~\cite{Morineau25}. We focus on \hto\ and \dto, archetypal spin ice materials with large susceptibilities and high-quality single crystals. Below approximately 2~K, these compounds enter the spin ice regime~\cite{Jaubert_book}, where the low-energy excitations are described as emergent magnetic monopoles whose complex dynamics are governed by correlated motion and kinetic constraints. Establishing how demagnetizing fields enter the experimentally measured fluctuation and response functions is therefore essential for a quantitatively controlled description of monopole dynamics. More broadly, our results demonstrate that boundary conditions enter fluctuation spectroscopy on equal footing with linear response, elevating sample geometry from a technical nuisance to a tunable parameter in the study of correlated spin dynamics.

The remainder of the paper is organized as follows. After detailing the experimental methods and sample geometries in Section~\ref{II}, we discuss the careful correction of ac susceptibility data for demagnetizing effects in Section~\ref{III}, and compare the corrected susceptibility with noise measurements for \hto\ in Section~\ref{IV}. The role of demagnetization on the noise spectra and their interpretation is addressed quantitatively in Section~\ref{V}. Section~\ref{VI} examines the impact of these effects on the key dynamical parameters and their temperature dependence. 
We show that correcting noise measurements for demagnetizing effects is necessary for an accurate description of the spin dynamics, and in particular that such corrections are essential for revealing the nature of monopole motion encoded in the high-frequency dependence of the noise spectrum.

\section{Experimental methods} \label{II}
Spectral noise density and ac susceptibility are measured in the temperature range from  4.2~K down to 600~mK using a custom-built SQUID magnetometer coupled to a dilution refrigerator. The specific sample is mounted in a first-order gradiometer configuration \cite{Morineau25}, which allows the detection of the magnetic field component produced by the sample along the coil axis.

\begin{table}[h!]
\caption{Summary of the sample configurations with their estimated demagnetization factor from Chen's method \cite{Chen05, Chen06}, assuming $\chi_{\rm int} \to 0$ (which is equivalent to  Arahoni's method in the case of parallelepipeds \cite{Aharoni98}) and $\chi_{\rm int}=0.4$~emu.cm$^{-3}$ (typical value for \hto\ at 1 K) and from the experimental method presented in Sec~\ref{III} using the \hto\ sphere sample as a reference. $N_{\rm DTO}$ is not accurately determined because the sample shape is not a perfect parallelepiped. The experimental values depend on the sphere measurements and are thus prone to errors which depend on the accuracy of the absolute values of the measured sphere susceptibility.}
\begin{tabular}{*6{Sc}}
\hline \hline
Config & Sphere & Para & Perp & Cylinder & DTO \\  \hline
\raisebox{15pt}{Shape} &
\includegraphics[width=1.2cm]{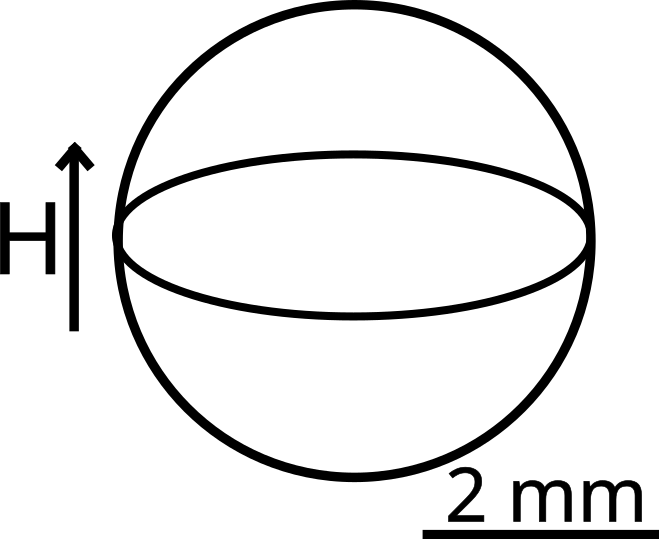} &
\includegraphics[height=1.15cm]{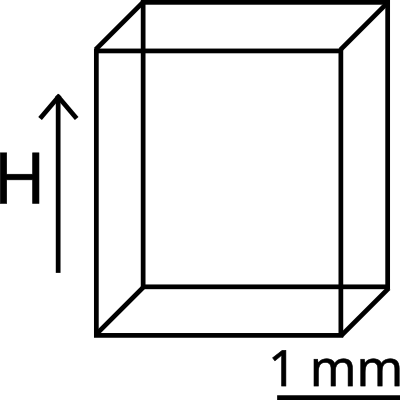} &
\includegraphics[height=0.8cm]{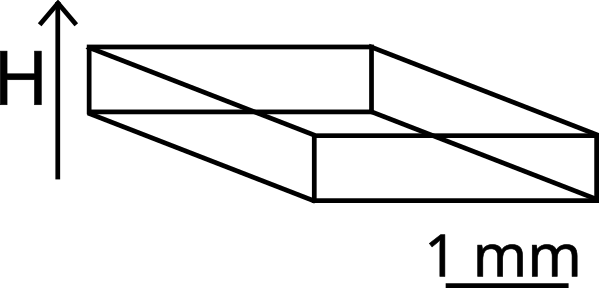} &
\includegraphics[height=1.4cm]{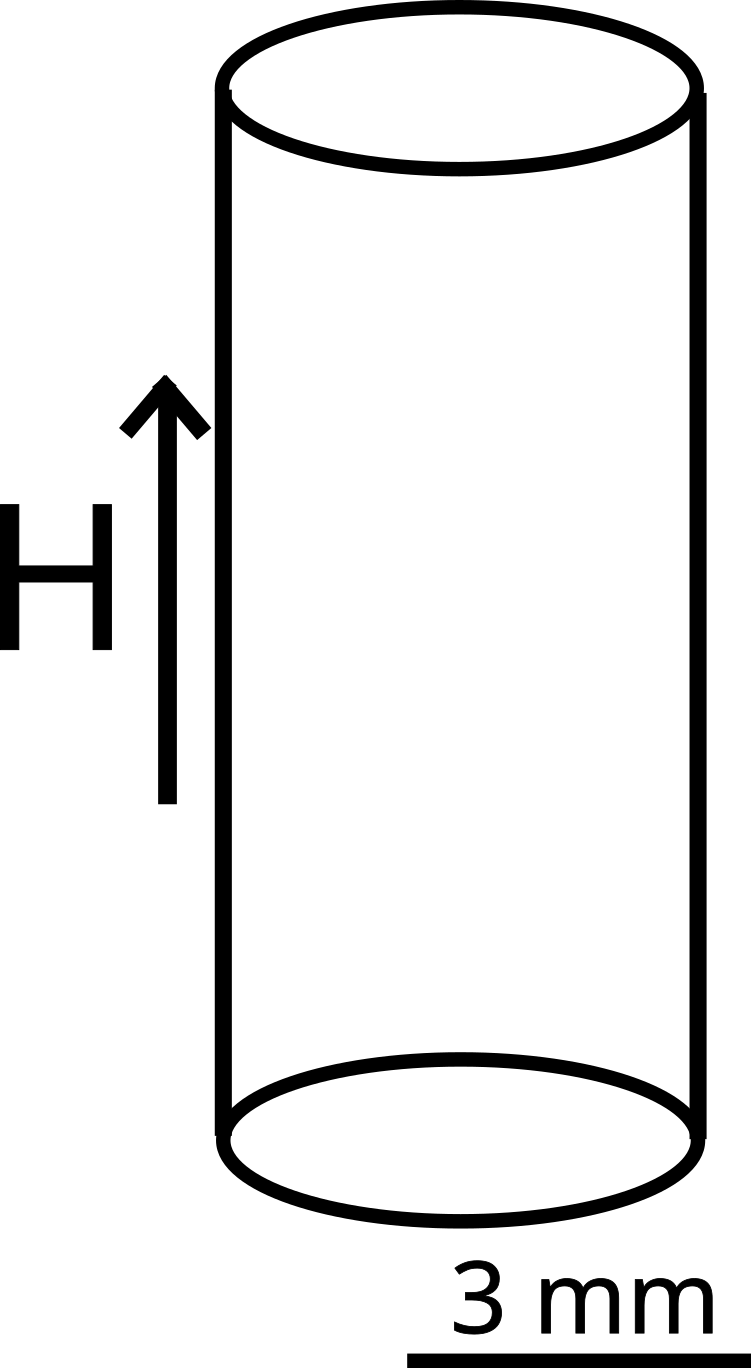} &
\includegraphics[height=1.4cm]{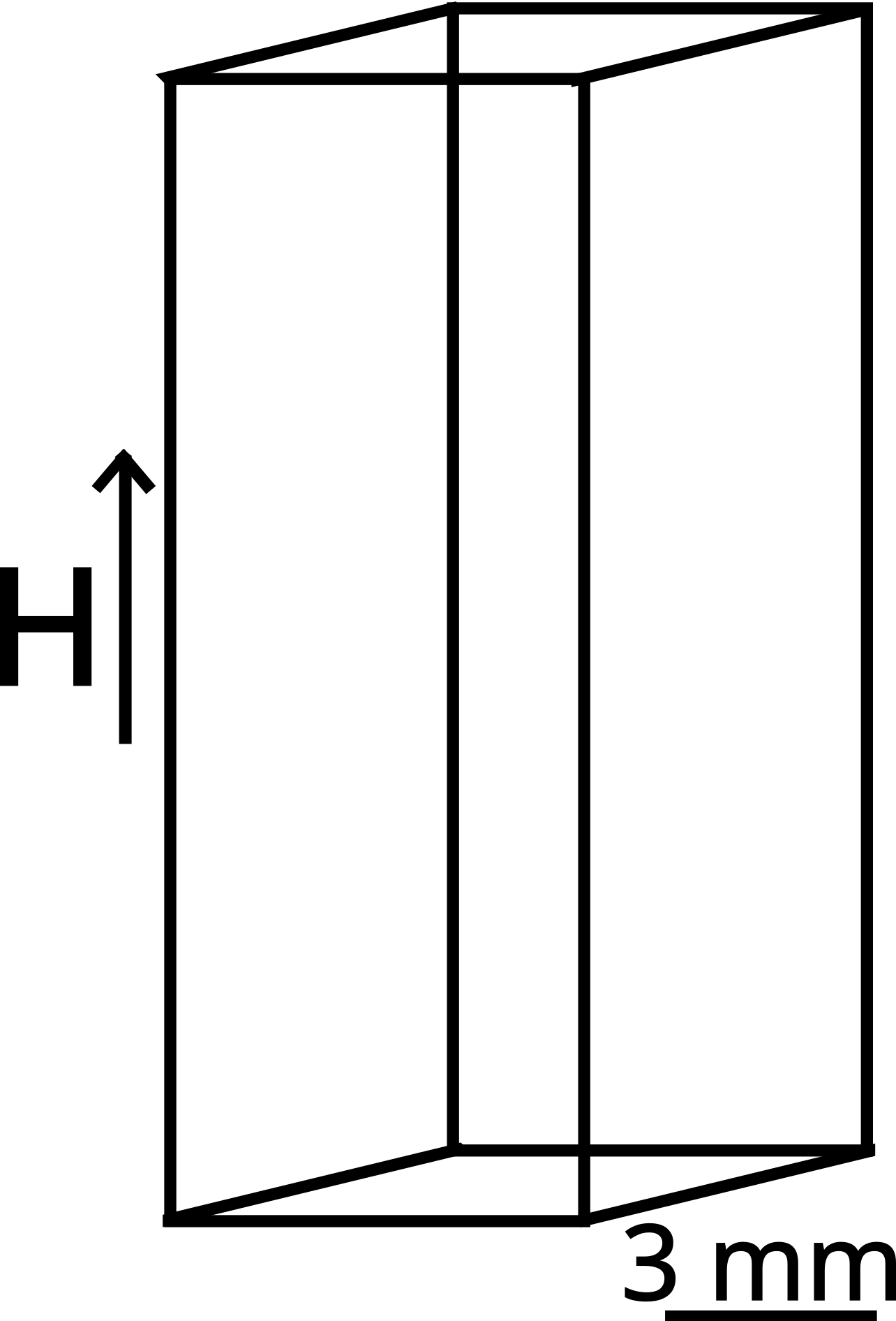} \\ 
$N$ label & $N_{\rm s}$  & $N_\parallel$  & $N_\perp$  & $N_{\rm c}$  & $N_{\rm DTO}$  \\ 
$\dfrac{N({ \scriptstyle \chi_{\rm int} \to 0} )}{4\pi}$ & $1/3$ & 0.163  & 0.646  & 0.214  & 0.131  \\ 
$\dfrac{N\!({\scriptstyle \chi_{\rm int}=0.4} )}{4\pi}$ & $1/3$ & 0.138  & 0.597  & 0.206  & 0.105  \\ 
$\dfrac{N_{\rm exp}}{4\pi}$ & $1/3$  & 0.093(2)  & 0.537(3)  & 0.177(2)  & -  \\ \hline \hline
\end{tabular}
\label{table_sample}
\end{table}

 Experiments were performed on two canonical spin ice materials, \hto\ and \dto. For \dto\ a single parallelepiped sample of isotopically enriched $^{162}$\dto\ with dimensions $3.7 \times 5.1 \times 14 $~mm$^3$  was used \cite{Giblin18, Fennell04}. For \hto\, three samples of different geometries were studied: a sphere of diameter 4 mm \cite{Bovo13}, a parallelepiped of dimensions $2.07 \times 2.41 \times 0.57$~mm$^3$, and a nearly cylindrical sample of diameter 4 mm and length 6.55~mm \cite{Morineau25}, the latter measured along its long axis. The parallelepiped \hto\ sample was measured along two distinct directions: one parallel to its longest dimension and one parallel to its shortest (i.e. perpendicular to the flat surface of the sample). Ac susceptibility measurements were performed by applying an alternating field of $H_{\rm ac} = 50~\mu$Oe  along the same direction as the field sensed by the gradiometer. The five sample configurations, along with the corresponding applied and detected field directions, are summarized in Table~\ref{table_sample}.

In the following, cgs units are used to express the ac susceptibility and noise data. However $N$ is more naturally expressed in SI units, where the sum over all direction equals 1. To make the conversion explicit, we write the $4\pi$ factor when reporting the $N$ values. 
 
\section{Sample shape effect on the susceptibility} \label{III}
Within the framework of linear response theory, the fluctuation-dissipation theorem relates equilibrium fluctuations quantified by the spectral noise density $S(f) = \langle M^2(f) \rangle$, where $M$ is the magnetization, to dissipation, quantified by the imaginary part of the ac susceptibility $\chi''(f)$, through the FDR:
 \begin{equation}
S(f)=\frac{2k_{\rm B}T}{\pi V} \frac{\chi''(f)}{f} 
\label{eq_FDT1}
\end{equation}
where $V$ is the sample volume, $T$ the temperature, and $k_{\rm B}$ the Boltzmann constant. By examining the influence of the demagnetization factor $N$ on the measured ac susceptibility, and computing the corresponding noise spectra via the FDR for samples of different geometries, the role of $N$ can be directly and quantitatively assessed. We therefore begin by analyzing the effect of $N$ on the ac susceptibility.

The measured ac susceptibility $\chi_{\rm meas} = \chi_{\rm meas}' - i\chi_{\rm meas}''$ is obtained from the in-phase and out-of-phase magnetization response to an applied alternating magnetic field ${\bf H}_{\rm a}$. For a uniformly magnetized sample of ellipsoidal geometry, the applied field is partially screened by a demagnetizing field ${\bf H}_{\rm d} = -[N]{\bf M}$, where $[N]$ is the demagnetizing tensor and ${\bf M}$ is the sample magnetization. The total internal field acting on the sample is then ${\bf H}_{\rm i} ={\bf H}_{\rm a} +{\bf H}_{\rm d}$. In our experimental configuration, the magnetization is measured along the direction of the applied field; consequently, only the scalar component of $[N]$ parallel to ${\bf H}_{\rm a}$ is relevant, and the internal field simplifies to $H_{\rm i} = H_{\rm a} - NM$. The intrinsic susceptibility $\chi_{\rm int} = dM/dH_{\rm i}$ is a shape independent material property, related to the measured susceptibility $\chi_{\rm meas} = dM/dH_{\rm a}$ by:
\begin{equation}
\chi_{\rm int}=\frac{\chi_{\rm meas}}{1-N\chi_{\rm meas}}
\end{equation}

Decomposing $\chi_{\rm meas} = \chi_{\rm meas}' - i\chi_{\rm meas}''$ into its real and imaginary parts yields:
\begin{equation}
\begin{cases}
\vspace{0.2cm}
\chi'_{\rm int} = \dfrac{\chi'_{\rm meas}-N(\chi'^2_{\rm meas}+\chi''^2_{\rm meas})}{(1-N\chi'_{\rm meas})^2+(N\chi''_{\rm meas})^2} \\
\chi''_{\rm int} = \dfrac{\chi''_{\rm meas}}{(1-N\chi'_{\rm meas})^2+(N\chi''_{\rm meas})^2} 
\end{cases}
 \label{chi2_demag}
\end{equation}

Equation~\ref{chi2_demag} implies that determining $\chi_{\rm int}'$ or $\chi_{\rm int}''$ requires knowledge of both $\chi_{\rm meas}'$ and $\chi_{\rm meas}''$. It follows that the intrinsic noise spectral density $S_{\rm int}(f)$ cannot be obtained from the measured noise spectral density $S_{\rm meas}(f)$ alone, since, by Equation \ref{eq_FDT1}, $S_{\rm meas}$ is related solely to $\chi_{\rm meas}''$. One approach to correcting $S_{\rm meas}(f)$ is to reconstruct $\chi_{\rm meas}'(f)$ via the Kramers--Kronig relations~\cite{Kronig1926}. However, this method requires integration over the entire frequency range where $\chi_{\rm meas}''$ is non-zero. For relaxation processes such as those in spin ice materials, the dynamical response spans a broad frequency range and is strongly temperature dependent~\cite{Matsuhira11}. In practice, this approach is therefore not applicable given the SQUID noise floor and the finite frequency bandwidth of the experiment. In Appendix \ref{Kramers-Kronig}, we illustrate the limits of this method with examples in spin ice at two different temperatures. 

A further challenge in correcting for demagnetization effects is the precise determination of $N$ for samples whose geometry is not always regular or well-defined. Except for ellipsoidal samples \cite{Osborn45}, the magnetization is inherently non-uniform, and an effective demagnetizing factor $N$ must be defined through an appropriate spatial average of ${\bf H}_{\rm d}$. An important consequence is that $N$ depends on the value of the susceptibility, and is therefore not constant as a function of temperature. The case of rectangular prisms has been studied extensively due to its experimental relevance, and an analytical expression for $N$ exists in the limit $\chi_{\rm int} \to 0$ that is widely used \cite{Aharoni98}. However, demagnetization effects are most significant when the susceptibility is large, requiring more refined estimates to accurately correct the experimental data. Micromagnetic calculations can be employed to estimate $N$ across the full susceptibility range within a continuum approximation, and tabulated values of $N$ are available for cylinders and rectangular prisms \cite{Chen05, Chen06}. More recently, it was shown using an iterative method and Monte Carlo simulations that accounting for the discrete nature of the magnetic moments and their crystallographic symmetry is necessary for a precise determination of $N$ \cite{Twengstrom17, Twengstrom20}. However, Ref.~\onlinecite{Twengstrom17} demonstrated that in the specific case of spin ice, corrections to the micromagnetic approach are small. We therefore adopt the tabulated values of Refs.~\onlinecite{Chen05, Chen06} for simplicity.

\begin{figure}[h!]
\includegraphics[width=8cm]{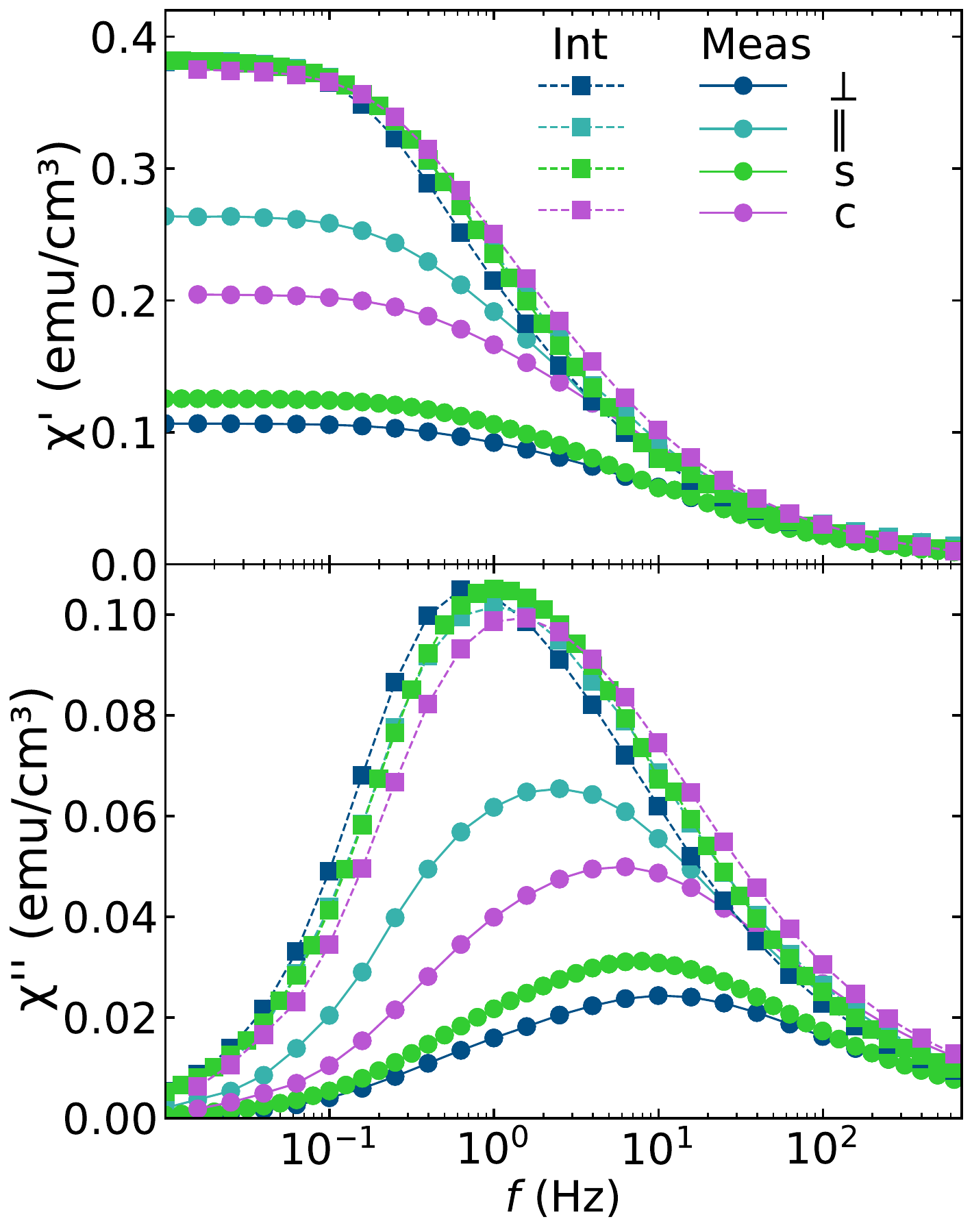}
\caption{\label{figXacmes}Ac susceptibility $\chi'$ and $\chi''$ vs $f$ measured at 900 mK for four different \hto\ sample configurations. The dots show the measured susceptibility $\chi_{\rm meas}$ and the squares the susceptibility corrected for  demagnetizing factor $N$ (intrinsic susceptibility $\chi_{\rm int}$). The $N_{\rm exp}$ factors used for the correction are : $N_{\perp} = 0.537(3)\times4\pi$, $N_{\parallel} = 0.093(2)\times4\pi$, $N_{\rm s} = 1/3 \times4\pi$ and $N_{\rm c} = 0.177(2)\times4\pi$. Lines are guides to the eye. }
\end{figure}

Beyond uncertainties in the definition of the sample geometry, a practical difficulty in applying these theoretical approaches to our measurements arises from the pickup coil geometry, which differs substantially from that of a conventional magnetometer. This is because the filling factor is optimized to maximize the signal, so that depending on the sample size, the measured response does not necessarily reflect the spatially averaged demagnetizing factor and may instead be sensitive to its non-uniform character. For this reason, in \hto\ we use the spherical sample for which $N$ is temperature-independent and equal to $1/3\times4\pi$ as a reference to determine $\chi_{\rm int}$ and to estimate effective values of the demagnetization factor $N_{\rm exp}$ for the other sample geometries \cite{Bovo13}.

Figure~\ref{figXacmes} shows the real and imaginary parts of the measured susceptibility ($\chi_{\rm meas}'$ and $\chi_{\rm meas}''$) and the corresponding demagnetization-corrected susceptibility ($\chi_{\rm int}'$ and $\chi_{\rm int}''$) for the four \hto\ sample configurations at 900~mK illustrated in Table \ref{table_sample}. The susceptibility of the spherical sample was first corrected using the known demagnetizing factor $N_{\rm s}=1/3\times4\pi$. The demagnetizing factors, $N$, for the remaining sample configurations were then fitted by minimizing the difference between their intrinsic susceptibilities and that of the spherical sample.
The resulting experimental demagnetizing factors are: $N_{\perp} = 0.537(3)\times4\pi$, $N_{\parallel} = 0.093(2)\times4\pi$ and $N_{\rm c} = 0.177(2)\times4\pi$. The close overlap of the intrinsic susceptibilities across all configurations confirms that the three samples share the same intrinsic dynamical properties, and that the differences observed in the measured susceptibilities arise predominantly from demagnetization effects. Further support for this conclusion comes from the high-frequency regime (above $10^2$~Hz), where $\chi'$ approaches zero and demagnetization corrections become negligible. In this limit, all susceptibilities naturally converge, independently of sample geometry.

For comparison, values of $N$ calculated in the limits $\chi_{\rm int} \to 0$ and $\chi_{\rm int} = 0.4$~emu~cm$^{-3}$ \cite{Aharoni98, Chen05, Chen06} are listed in Table~\ref{table_sample}. The experimentally determined values fall within the same range but are systematically lower than the calculated ones. Since our samples are large relative to the pickup coil, this discrepancy likely reflects the fact that our measurement preferentially probes the local demagnetizing factor near the sample center, which is expected to be smaller than the spatial average \cite{Chen05, Chen06}.  

\begin{figure}[h!]
\begin{center}
\includegraphics[width=7.5cm]{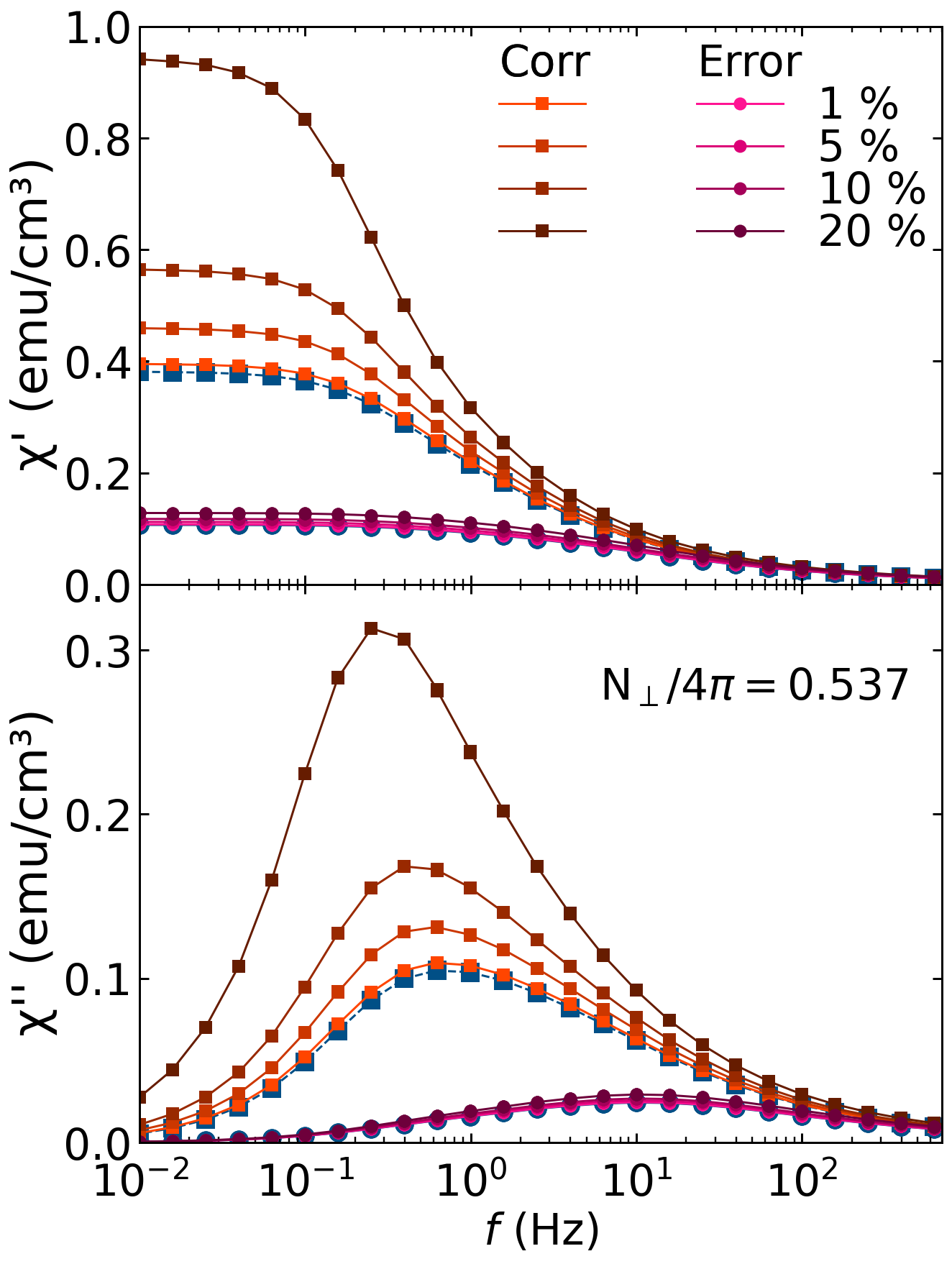}
\end{center}
\caption{\label{cube_para_error}
Real ($\chi'$) and imaginary ($\chi''$) parts of the ac susceptibility as a function of frequency, measured at 900~mK for the \hto\ parallelepiped sample in the perpendicular configuration ($N_{\perp} = 0.537(3)\times 4\pi$). Dots show the measured susceptibility and squares the demagnetization-corrected intrinsic susceptibility. The reference measured and intrinsic susceptibilities (with no artificial error) are shown in blue. To illustrate error propagation, the measured susceptibility is artificially scaled by +1, +5, +10, and +20\% (shades of pink); the corresponding intrinsic susceptibilities obtained after demagnetization correction are shown in shades of orange. The strong deviation of the orange curves from the blue reference demonstrates the amplification of measurement errors by the demagnetization correction.}
\end{figure}

It is worth noting that for large $N$, small errors in the measured susceptibility can be significantly amplified after demagnetization correction. This is particularly relevant for spin ice materials such as \hto\ and \dto\, which exhibit large values of $\chi'$ due to ferromagnetic interactions.
To illustrate this sensitivity, we examine how errors in $\chi_{\rm meas}$ propagate through the demagnetization correction for the two sample configurations with the largest $N$ values measured: the parallelepiped in the perpendicular orientation and the sphere.
Figure~\ref{cube_para_error} shows, for the \hto\ configuration with $N_{\perp} = 0.537(3)\times 4\pi$, the effect of introducing artificial errors of 1, 5, 10, and 20\%  in $\chi_{\rm meas}$ (shown in pink) on the corresponding corrected intrinsic susceptibility $\chi_{\rm int}$ (shown in orange). The original measured and intrinsic susceptibilities, with no artificial errors, are shown in blue (nearly coincident with the 1\% curve). This analysis demonstrates that errors are substantially amplified by the demagnetization correction. Quantifying this effect through the magnitude of $\chi_{\rm int}'(f \to 0)$ after correction, an initial error of 1\% propagates to 4.2\%, 5\% to 20\%, 10\% to 49\%, and 20\% to 152\%.

\begin{figure}[h!]
\begin{center}
\includegraphics[width=7.5cm]{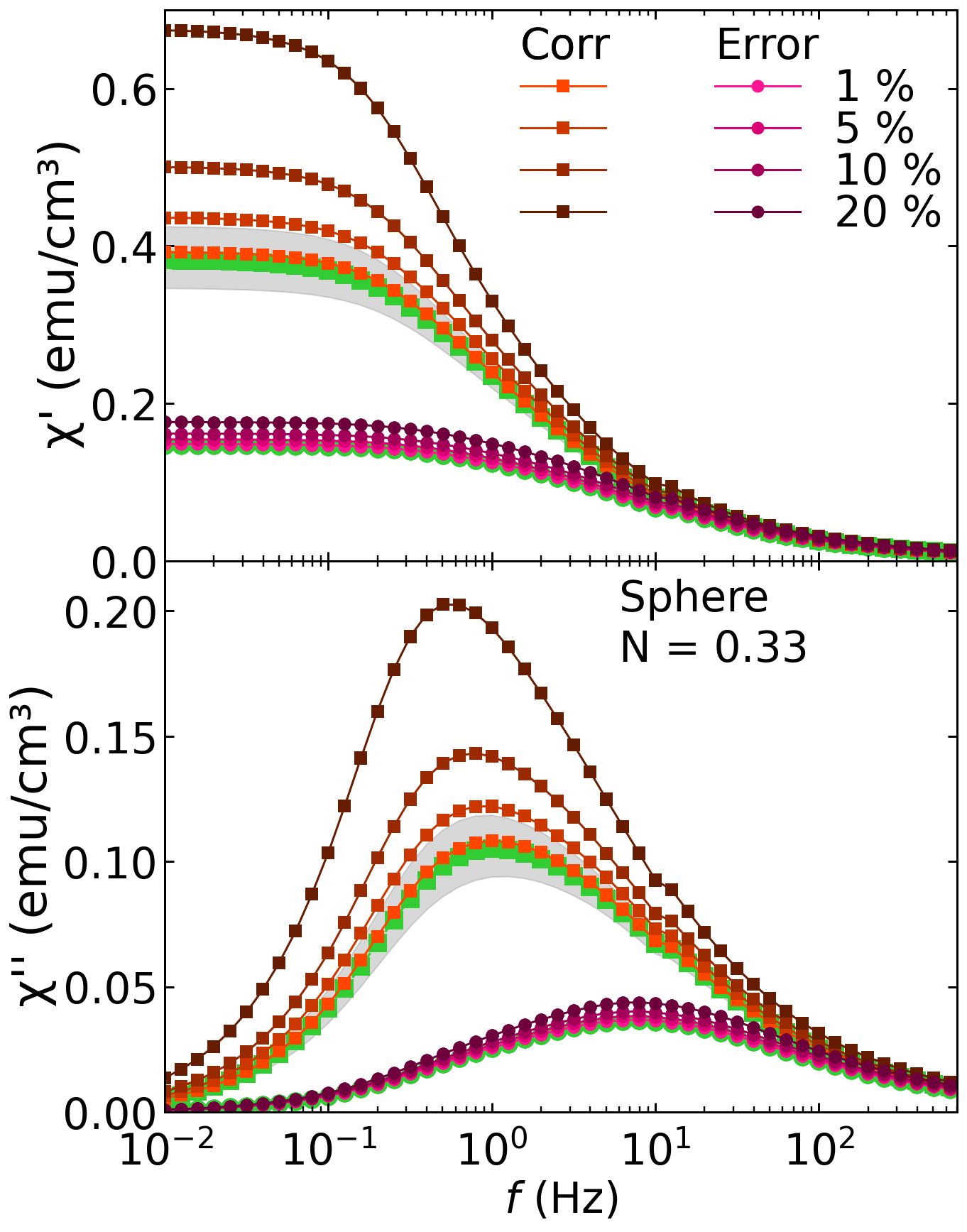}
\end{center}
\caption{\label{sphere_error} Real ($\chi'$) and imaginary ($\chi''$) parts of the ac susceptibility as a function of frequency, measured at 900~mK for the \hto\ spherical sample ($N_{\rm s} = 1/3\times 4\pi$). Dots show the measured susceptibility and squares the demagnetization-corrected intrinsic susceptibility. The reference measured and intrinsic susceptibilities (with no artificial error) are shown in green. To illustrate error propagation, the measured susceptibility is artificially scaled by +1, +5, +10, and +20\% (shades of pink); the corresponding intrinsic susceptibilities obtained after demagnetization correction are shown in shades of orange. The grey bandwidth indicates the uncertainty resulting from our calibration error for the sphere. Compared to Figure~\ref{cube_para_error}, the error amplification is reduced, consistent with the smaller value of $N_{\rm s}$, though still remaining substantial.}
\end{figure}

A similar analysis for the \hto\ sphere configuration is shown in Figure~\ref{sphere_error}. Starting from initial artificial errors of 1, 5, 10, and 20\%, the demagnetization correction yields final errors of 3.2, 14.5, 32, and 77\%, respectively. The error amplification is less pronounced than in the parallelepiped perpendicular geometry, as expected from the smaller value of $N$. Nevertheless, $N_{\rm s} = 1/3\times 4\pi$ remains a substantial demagnetizing factor, and this demonstrates that precise absolute calibration of $\chi_{\rm meas}$ is essential to obtain accurate values of $\chi_{\rm int}$.
Furthermore, in the correction procedure applied in Figure~\ref{figXacmes}, any errors in the absolute value of $\chi_{\rm meas}$ for the sphere arising for instance from calibration uncertainties would also affect the derived effective demagnetizing factors for the remaining sample configurations.

As a consequence, while the spherical sample offers an appealing reference geometry owing to its well defined and temperature-independent $N$, uncertainties in the absolute calibration of $\chi_{\rm meas}$ significantly impact the accuracy of $\chi_{\rm int}$, precisely because $N_{\rm s}$ is not negligibly small. It therefore appears preferable to minimize $N$ rather than to rely on its precise knowledge. In principle, a sample with a small demagnetizing factor such as a long ellipsoid would provide a more robust route to $\chi_{\rm int}$. In practice, however, such geometries are difficult to achieve experimentally and yield a poor filling factor, making them suboptimal for noise measurements.

In summary, the intrinsic susceptibility of our samples can be estimated using a spherical sample as a reference, and the demagnetizing factors derived for the various \hto\ configurations by this method are taken as reliable inputs for the subsequent analysis. We now proceed to assess their impact on the noise measurements.
Since noise and ac susceptibility measurements are performed on the same sample in the same experimental run, the measurement geometry, orientation, position, and filling factor  is identical for both, enabling a direct quantitative comparison.

\section{The effect of sample geometry on the noise}
\subsection{Experimental observations} \label{IV}

With the demagnetization corrections to the ac susceptibility established, the fluctuation-dissipation theorem (Equation~\ref{eq_FDT1}) provides a natural framework for estimating their effect on the noise spectra. To distinguish the power spectral density $S(f)$ obtained directly from noise measurements from its counterpart $D(f)$ derived via the FDR from the ac susceptibility, we define: 
 \begin{gather}
S(f)=D(f)  \label{eq_FDT} \\
 \textrm{\quad where \ } S(f)=\overline{M^2(f)} \textrm{\quad and \ } D(f)=\frac{2k_{\rm B}T}{\pi V} \frac{\chi''(f)}{f} \nonumber
\end{gather}

\begin{figure}[h!]
\includegraphics[width=8.8cm]{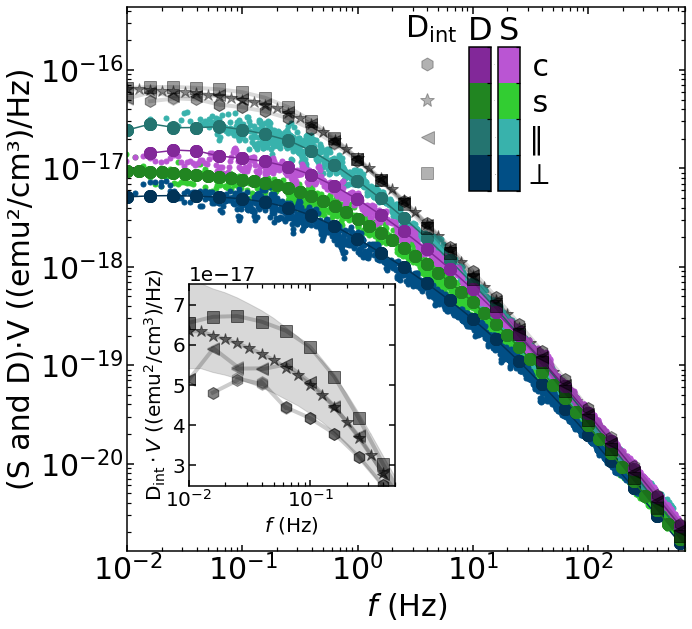}
\caption{\label{fignoisemes} FDR plot multiplied by volume for four different sample configurations of \hto\ on a logarithmic scale: $S_{\rm meas} \times V$ (small dots) and $D_{\rm meas}\times V$ (big dots) as a function of frequency $f$ measured at 900 mK. The gray symbols represent $D_{\rm int} \times V$, which was corrected for demagnetizing effects with the $N$ factors: $N_{\perp} = 0.537(3)\times4\pi$, $N_{\parallel}= 0.093(2)\times4\pi$, $N_{ \rm s} = 1/3\times4\pi$ and $N_{ \rm c} = 0.177(2)\times4\pi$. Inset: Magnification of the low frequency region of $D_{\rm int}$, with a vertical linear scale. The light grey bandwidth indicates the error estimated on the intrinsic value from the uncertainty on the sphere susceptibility.  Lines are guides to the eye.}
\end{figure}

In our previous experimental work~\cite{Morineau25}, we demonstrated that, at all frequencies and for temperatures above 400~mK, $S_{\rm meas}(f)$ agrees with $D(f)$ when the latter is computed from the measured out-of-phase susceptibility $\chi_{\rm meas}''(f)$. As discussed in Section~\ref{III}, this agreement directly implies that $S_{\rm meas}(f)$ is a sample and shape dependent quantity, just as $\chi_{\rm meas}(f)$ is.

This is illustrated in Figure~\ref{fignoisemes}, which shows the measured noise spectra $S_{\rm meas}(f)$ for the four \hto\ sample configurations at 900~mK, in excellent agreement with $D_{\rm meas}(f)$ derived from the measured susceptibility via the FDR.
To facilitate comparison across configurations, all spectra are normalized by the sample volume $V$, yielding the volume-normalized quantities $VS(f)$ and $VD(f)$.

As with the ac susceptibility, the measured noise is clearly shape dependent, with significant variation in the low-frequency noise plateau value $S_{0,{\rm meas}} = S_{\rm meas}(f \to 0)$ across configurations: the ratio between the plateau values of the two parallelepiped configurations reaches a factor of 6. The intrinsic noise spectrum $S_{\rm int}(f)$ is obtained by computing $D_{\rm int}(f)$ from the intrinsic susceptibility determined in Section~\ref{III}, and is shown in black in Figure~\ref{fignoisemes} for all four configurations. The corrected spectra converge to a common plateau value of $6 \times 10^{-17}$~emu$^2$.cm$^{-3}$.Hz$^{-1}$, which is nearly one order of magnitude larger than the plateau value measured directly in the perpendicular configuration ($N_\perp=0.537(3)\times 4\pi$). In addition, the characteristic relaxation times are systematically shifted to lower frequencies by the demagnetization correction. Taken together, these observations underscore the importance of accounting for sample geometry in order to quantitatively access the intrinsic spin dynamics.

\subsection{Quantitative effects of $N$ on the noise} \label{V}
These observations are consistent with the findings of Ref.~\onlinecite{Bramwell25}, which establishes that the noise autocorrelation function $S_{\rm meas}(t) = \langle M(0)M(t) \rangle$ depends on sample shape through the canonical average $\langle \cdots \rangle$, as does its Fourier transform $S_{\rm meas}(f)$. It is further shown that the noise power spectral density (PSD) can be analytically corrected for demagnetization effects in specific cases such as exponential or pink noise relaxation but that no general correction exists. For monotonic relaxation characterized by a mean relaxation time $\tau$, the effect of demagnetization can nevertheless be quantified through the ratios of measured to intrinsic values of the zero-frequency noise $S_0$ and the relaxation time $\tau$:
\begin{align}
\frac{S_{\rm 0,meas}}{S_{\rm 0,int}}&= \frac{1}{(1+N \chi_{\rm 0,int})^2}\label{eq_ratio}\\
\frac{\tau_{\rm meas}}{\tau_{\rm int}}&= \frac{1}{1+N\chi_{\rm 0,int}}\label{eq_ratio2}
\end{align}
where $\chi_{\rm 0, int}$ is the static susceptibility. 

In the following, we quantitatively assess the impact of demagnetization on the measured noise spectra in spin ice materials. These systems exhibit complex dynamics involving distributions of relaxation times. The empirical PSD associated with such dynamics can be described by \cite{Dusad19, Samarakoon22b}:
\begin{align}
S(f)&=\frac{S_0}{1+(2\pi f \tau)^{\alpha}} \label{Davidson_Cole_noise}
\end{align}
with $S_0$ the noise value at zero frequency, $\tau$ the characteristic relaxation time and $\alpha$ the anomalous exponent. 

Since no simple analytical expression for the susceptibility can be derived from Equation~\ref{Davidson_Cole_noise}, we instead consider the Davidson-Cole model of the susceptibility to estimate the effect of $N$ on both $S(f)$ and $\tau$ across the full frequency range: 
\begin{align}
 \chi(f)&=\chi_{\rm S}+\dfrac{\chi_{\rm T}-\chi_{\rm S}}{(1+i2\pi f \tau)^{\beta}}\label{Davidson_Cole}
\end{align}
where $\chi_{\rm S}$ and $\chi_{\rm T}$ are the adiabatic and isothermal susceptibilities, respectively, $\beta \in [0,1]$ is a stretching exponent, and we set $\chi_{\rm S} = 0$ for simplicity.
This model yields results close to though not strictly equivalent to Equation~\ref{Davidson_Cole_noise}.

In \hto\, two distinct relaxation processes have been identified \cite{Wang21, Morineau25}, requiring a sum of two Davidson-Cole contributions (Equation~\ref{Davidson_Cole}) or equivalently two terms of the form of Equation~\ref{Davidson_Cole_noise} to describe the dynamics. To avoid this additional complexity, we first focus on \dto\, which exhibits a single relaxation process.
The measured susceptibility at 900~mK is fitted using Equation~\ref{Davidson_Cole}. The fit is then corrected for demagnetization using $N_{\rm DTO} = 0.105\times 4\pi$, estimated from the method of Ref.~\onlinecite{Chen05} (see Table~\ref{table_sample}), to recover the intrinsic susceptibility. The latter is subsequently backward-modelled \footnote{Refers to the use of Equation~\ref{chi2_demag} in the reverse direction, expressing $\chi_\mathrm{meas}$ as a fonction of $\chi_\mathrm{int}$.}, giving the measured susceptibility for various $N$ from 0 to $4\pi$ corresponding to a set of different geometries. The intrinsic and measured noise spectra, $S_{\rm int}(f) = D_{\rm int}(f)$ and $S_{\rm meas}(f) = D_{\rm meas}(f)$, are then computed via the FDR (Equation~\ref{eq_FDT}) for each value of $N$.

\begin{figure}
\includegraphics[width=4.2cm]{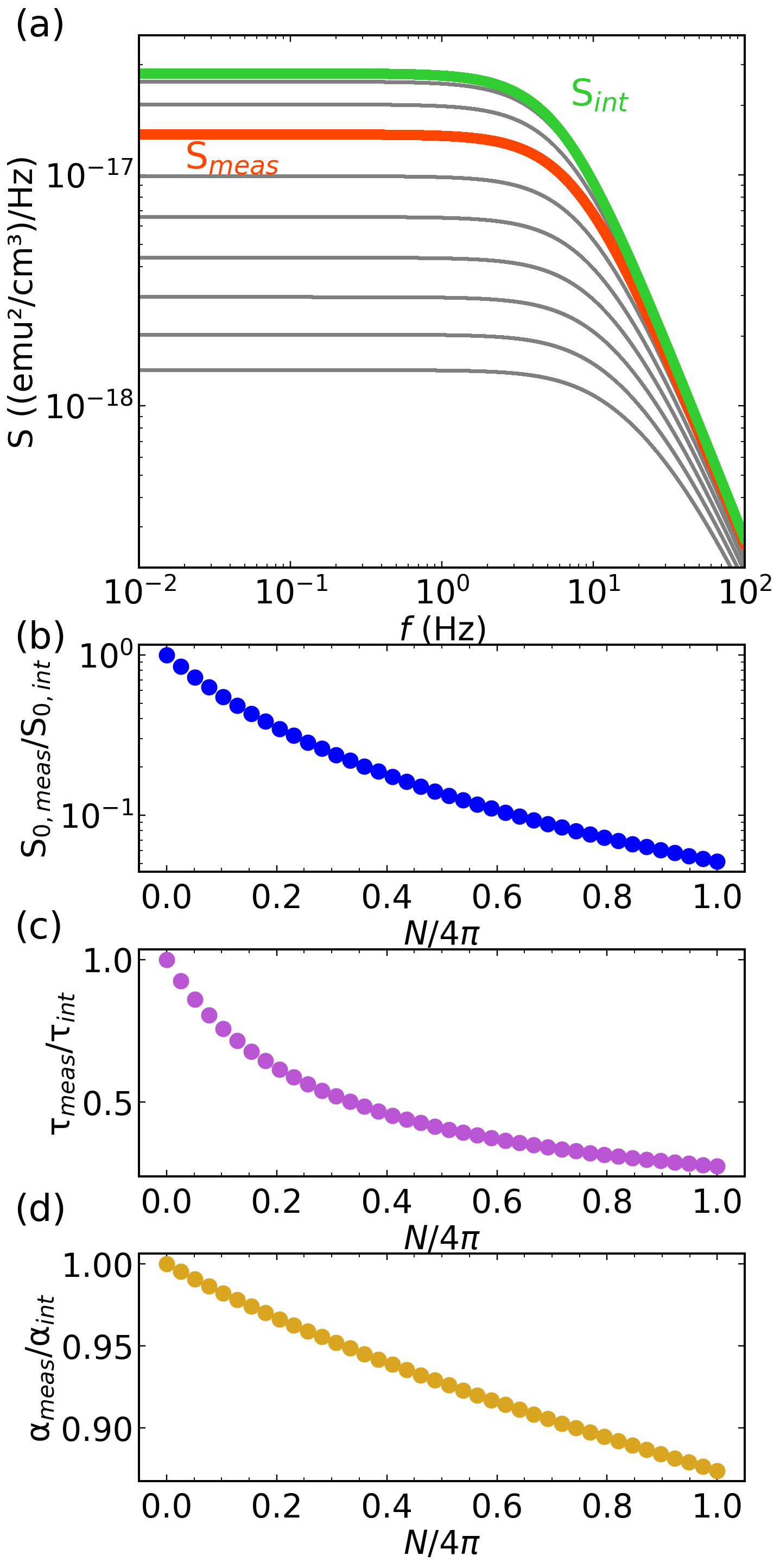}
\includegraphics[width=4.2cm]{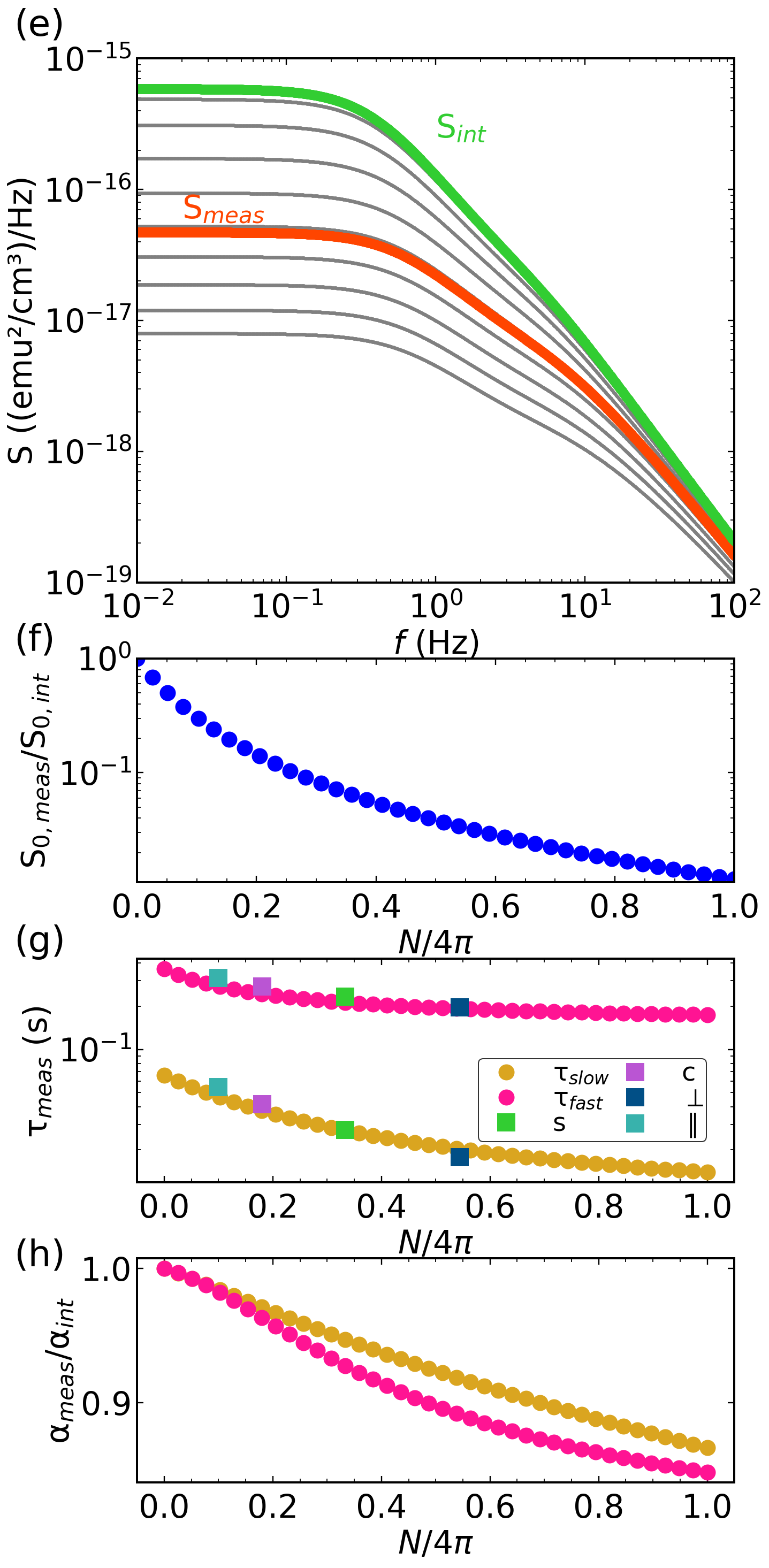}
\caption{\label{figXnoisemodel} Impact of the demagnetization factor on the noise and the fitted parameters in (a)-(d) \dto\  and  (e)-(h) \hto. (a)(e) $S$ vs $f$ for different $N$ values between 0 and $4\pi$ at 900 mK. In (a) the orange curve $S_{\rm meas}$ is calculated from the FDR using a Cole-Davidson function for the susceptibility with $\tau=0.022$~s, $\chi_T=0.20$~emu.cm$^{-3}$, $\beta=0.64$, obtained from the fit of the experimental data. In (e) a sum of two Cole-Davidson equations is used with $\tau_1=0.29$~s, $\beta_1=0.40$, $\chi_{\rm T1}=0.081$ emu.cm$^{-3}$, $\tau_2=0.026$ s, $\beta_2=0.43$, $\chi_{\rm T 2}=0.090$~emu.cm$^{-3}$. The green curves $S_{\rm int}$ are obtained by correcting the orange curves using (a) $N_{\rm DTO} = 0.105\times4\pi$ and (e) $N_{\rm s} = 1/3\times4\pi$. The grey curves correspond to the noise $S_{\rm meas}$ that is expected to be measured for samples with different values of $N$. (b)(f) Measured magnetic noise plateau $S_{\rm 0,meas}$ normalized by the intrinsic magnetic noise plateau $S_{\rm 0,int}$ as a function of $N$. (c) Relaxation time ratio as a function of $N$ obtained from the fit of the $S(f)$ curves in (a) using Equation \ref{Davidson_Cole_noise}. (g) Relaxation times $\tau_{\rm slow}$ (pink) and $\tau_{\rm fast}$ (gold) as a function of $N$ obtained from the fit of the $S(f)$ curves in (e) with a sum of two Equations \ref{Davidson_Cole_noise}. The squares show the times obtained from the different samples measured in this study. The intrinsic relaxation times are the values at $N=0$.  (d)(h) $\alpha_{\rm meas}/\alpha_{\rm int}$ ratios for different values of $N$ extracted from the same fitting procedure as the two other parameters. In (h) the fast and slow $\alpha$ ratios are shown in pink and gold respectively. 
 }
\end{figure}

Figure~\ref{figXnoisemodel}(a) shows the noise spectra $S(f)$ (grey curves) computed for values of $N$ ranging from 0 to $4\pi$. The spectrum obtained from the fit of the experimental susceptibility is highlighted in orange and serves as the reference, while the corresponding intrinsic noise $S_{\rm int}(f)$ is shown in green. This demonstrates that $N$ can dramatically affect $S_{\rm meas}(f)$, with variations exceeding one order of magnitude between $N = 0$ and $N = 4\pi$.
The zero-frequency parameters $S_{\rm 0,meas}$, $S_{\rm 0,int}$, and $\chi_{\rm 0,int}$ are extracted from the low-frequency limit of the respective curves. The characteristic relaxation times $\tau$ are obtained by fitting the noise spectra with Equation~\ref{Davidson_Cole_noise}, following the approach used in previous noise studies of spin ice. The obtained dependence of the $S_0$ and $\tau$ ratios on $N$ is shown in panels (b) and (c) of Figure~\ref{figXnoisemodel}. Both ratios are in excellent agreement with the analytical expressions of Equations~\ref{eq_ratio} and~\ref{eq_ratio2}.

We repeated this procedure for \hto, which exhibits two relaxation processes, using a sum of two Davidson-Cole contributions (Equation~\ref{Davidson_Cole}). The data obtained for the \hto\ sphere at 900~mK are fitted, and $N_{\rm s} = 1/3\times 4\pi$ is used to recover the intrinsic spectrum, which is then ``backward-modelled" for other values of $N$. The results are shown in the right panels (e--h) of Figure~\ref{figXnoisemodel}.
The ratio $S_{\rm 0,meas}/S_{\rm 0,int}$ follows Equation~\ref{eq_ratio} as a function of $N$ and $\chi_{\rm 0,int}$, and as in the \dto\  case, the resulting $S_0$ values span more than one order of magnitude.
The behavior of the $\tau$ ratios is more complex. The relaxation times extracted from the noise spectra decrease with increasing $N$, but the ratios deviate from the predicted $1/(1+N\chi_{\rm 0,int})$. This is likely due to the two relaxation processes not being well separated: if the two time distributions overlap, the fitted parameters are less accurate.
The relaxation times obtained by fitting the \hto\ noise data at 900~mK for the different sample configurations are also shown in Figure~\ref{figXnoisemodel}(g), and are consistent with the backward-modelling results of Figure~\ref{figXnoisemodel}(e). For the parallelepiped perpendicular configuration, which has the largest $N$, the slow and fast relaxation time ratios are $(\tau_{\rm meas}/\tau_{\rm int})_{\rm slow} = 0.266$ and $(\tau_{\rm meas}/\tau_{\rm int})_{\rm fast} = 0.414$, respectively, demonstrating that demagnetization-induced errors in $\tau$ can be substantial and that their correction is not straightforward.

One might expect that demagnetization corrections should not have a large effect on the exponent $\alpha$, since it governs the high-frequency behavior where demagnetization effects vanish. In practice, however, Figures~\ref{figXnoisemodel}(a) and (e) show that this high-frequency limit is not reached in either the single- or double-relaxation cases. As a consequence, the observed spectral slopes -- and hence the fitted exponents -- also depend on $N$ and tend to be underestimated. The ratios $\alpha_{\rm meas}/\alpha_{\rm int}$, obtained by fitting the noise spectra over our experimental frequency range, are shown in Figures~\ref{figXnoisemodel}(d) and (h). They decrease monotonically with $N$, reaching a value of approximately 0.85 at $N = 4\pi$. To minimize this bias, the fit of Equation~\ref{Davidson_Cole_noise} must be extended to sufficiently high frequencies. For instance, at $N = 4\pi$, a frequency range spanning more than four decades above $1/\tau$ is required to achieve an error of less than 5\% on $\alpha$.
In the \hto\ case, the behavior of the two $\alpha$ ratios is less clear-cut, owing primarily to the proximity of the two relaxation times, which both influence the slope in the intermediate frequency regime. As a result, extending the fit frequency range does not lead to a significant improvement.

These results confirm that the noise spectra and their fitted parameters are strongly influenced by $N$. This effect is particularly pronounced for the zero-frequency noise plateau $S_{\rm 0,meas}$, which follows the expected $1/(1+N\chi_{\rm 0,int})^2$ dependence. Accessing the intrinsic plateau value thus requires knowledge of the intrinsic low-frequency susceptibility $\chi_{\rm 0,int}$, which in turn necessitates an independent dc susceptibility measurement to perform the demagnetization correction.
As discussed in Section \ref{II} and Appendix \ref{Kramers-Kronig}, an alternative approach would be to model the ac susceptibility derived from a fit of the noise spectrum, thereby determining both its real and imaginary parts, and to reconstruct the demagnetization-corrected noise plateau without requiring an independent experimental determination of $\chi_{\rm 0,int}$.

In summary, fitting the noise spectra with phenomenological models allows the dependence of $S_0$, $\tau$, and $\alpha$ on $N$ to be quantitatively assessed. The ratio $S_{\rm 0,meas}/S_{\rm 0,int}$ extracted from the fits follows the expected analytical scaling~\cite{Bramwell25}. The behavior of the $\tau$ ratios is more nuanced and depends on the nature of the relaxation process -- in particular, presence of a single or two characteristic relaxation times, width of the distribution. Finally, the fitted exponent $\alpha$, which should be $N$-independent, can also be affected by demagnetization if the measurements do not extend to sufficiently high frequencies.
 
\section{Impact of $N$ on the dynamical properties of spin ice materials} \label{VI}
In practice, experiments typically involve a single sample geometry measured over a range of temperatures, rather than fixing $T$ and varying $N$. It is therefore instructive to estimate the impact of demagnetization across the full temperature range. We examine this effect for one sample of each spin ice material studied: (i) the spherical \hto\ sample, whose demagnetizing factor $N_{\rm s} = 1/3\times 4\pi$ is temperature-independent
; and (ii) the parallelepiped \dto\ sample, used in the absence of a spherical sample for this material, whose demagnetizing factor $N_{\rm DTO} = 0.105\times 4\pi$ is relatively small. For the latter geometry, $N$ is weakly temperature dependent increasing by approximately 10\% between base temperature and 4~K but is treated as constant here for simplicity. This analysis is intended to illustrate the magnitude of demagnetization effects and does not aim to provide precisely corrected parameter values.

\begin{figure}[h!]
\includegraphics[width=7.5cm]{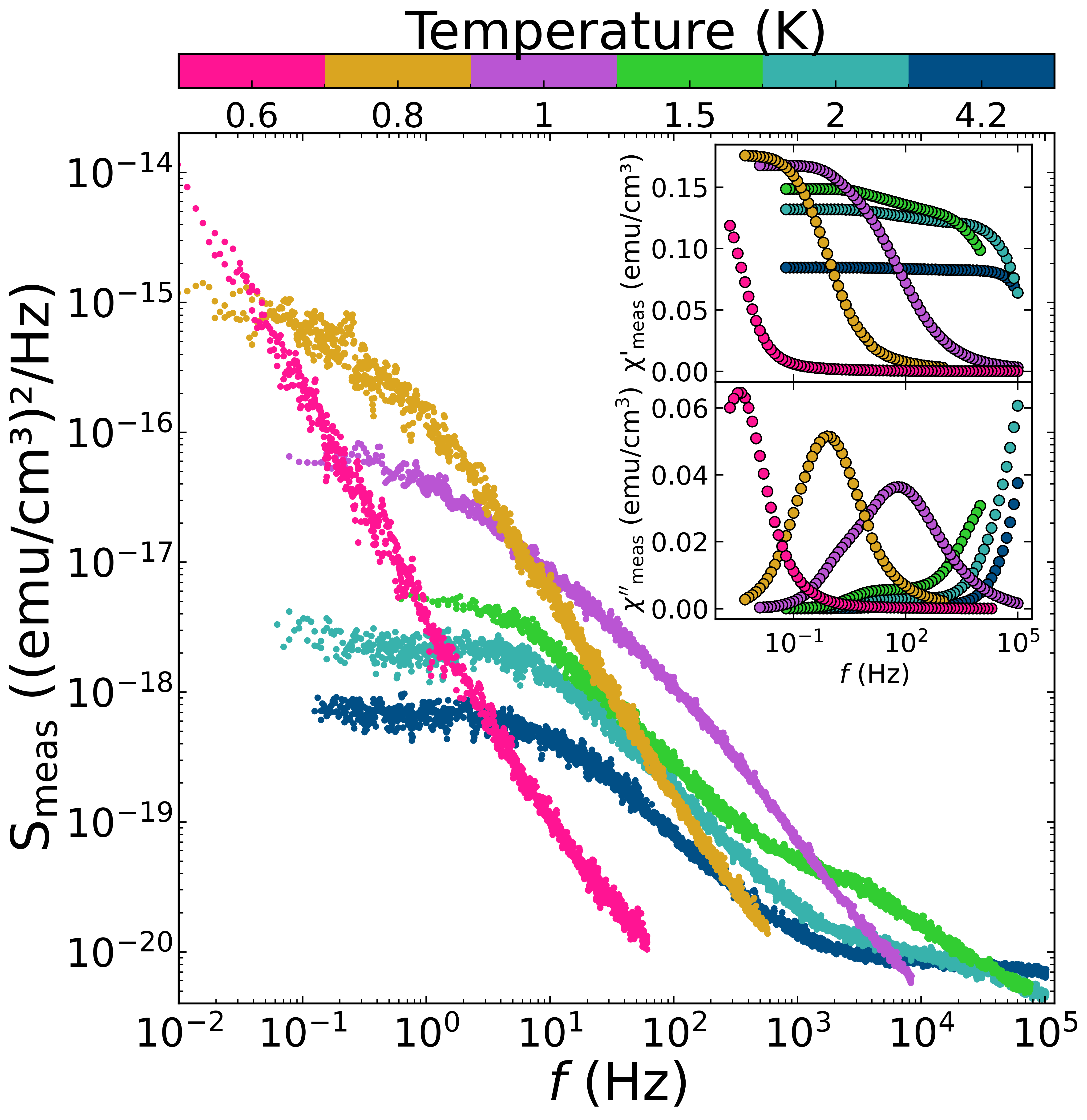}
\caption{\label{noise_sphere} Measured noise spectra $S_{\rm meas}(f)$ between 600 mK and 4.2 K for the \hto\ sphere sample. The inset shows the measured susceptibility vs frequency at the same temperatures, with $\chi'_{\rm meas}$ at the top and $\chi''_{\rm meas}$ at the bottom. }
\end{figure}

A representative selection of measured susceptibility and noise spectra across the 4.2--0.6~K temperature range is shown in Figure~\ref{noise_sphere} for the \hto\ sphere sample. The corresponding data for the \dto\ sample can be found in Ref.~\cite{Morineau25}. Below 500~mK, the system enters the spin freezing regime of spin ice \cite{Snyder04}, where the dynamics slow to timescales falling below the lowest accessible frequency of our experiment (0.001~Hz). In this regime, the susceptibility response is strongly suppressed and demagnetization corrections become negligible \cite{Morineau25}; it is therefore not considered further here.

To extract the characteristic dynamical parameters for both materials, we follow the procedure below. (i) The intrinsic noise spectra $D_{\rm int}(f)$ are obtained from the measured susceptibility corrected for demagnetization effects using Equations~\ref{chi2_demag} and \ref{eq_FDT}. (ii) Both $D_{\rm int}(f)$ and the measured noise spectra $S_{\rm meas}(f)$ are fitted to Equation~\ref{Davidson_Cole_noise} for temperatures between 600~mK and 4.2~K, where demagnetization effects are relevant. A single relaxation time model is used for \dto, while a two relaxation time model is used for \hto, as described in Section~\ref{V}. (iii) The impact of the demagnetization correction is quantified through the ratios $S_{\rm 0,meas}/S_{\rm 0,int}$, $\tau_{\rm meas}/\tau_{\rm int}$, and $\alpha_{\rm meas}/\alpha_{\rm int}$, evaluated across the full temperature range. At temperatures below 800~mK for \hto\ and 600~mK for \dto, the noise plateau falls outside the accessible frequency window, rendering the fitted parameters unreliable; these temperature ranges are therefore excluded from the subsequent analysis.

\begin{figure}[b]
\includegraphics[width=4.2cm]{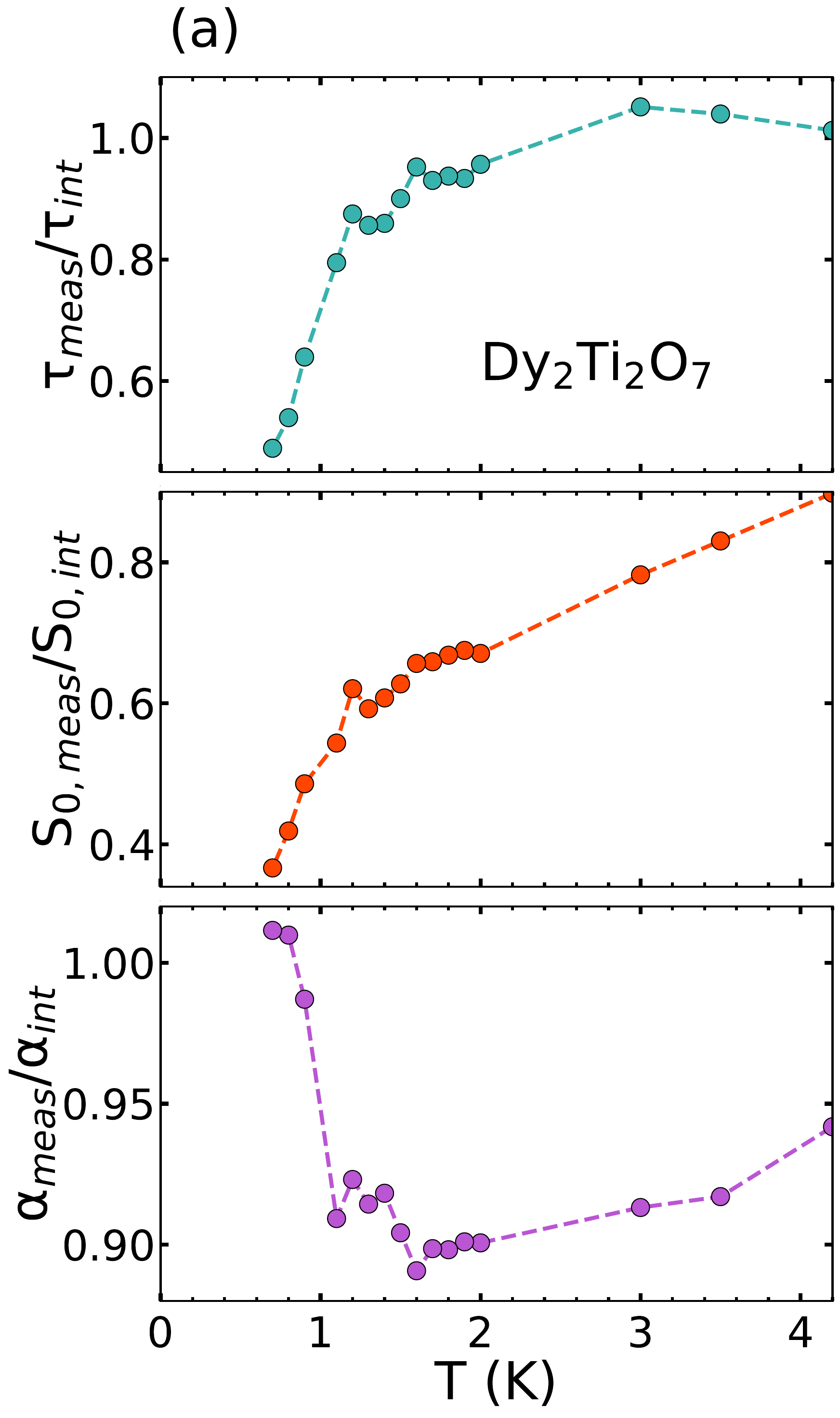}
\includegraphics[width=4.1cm]{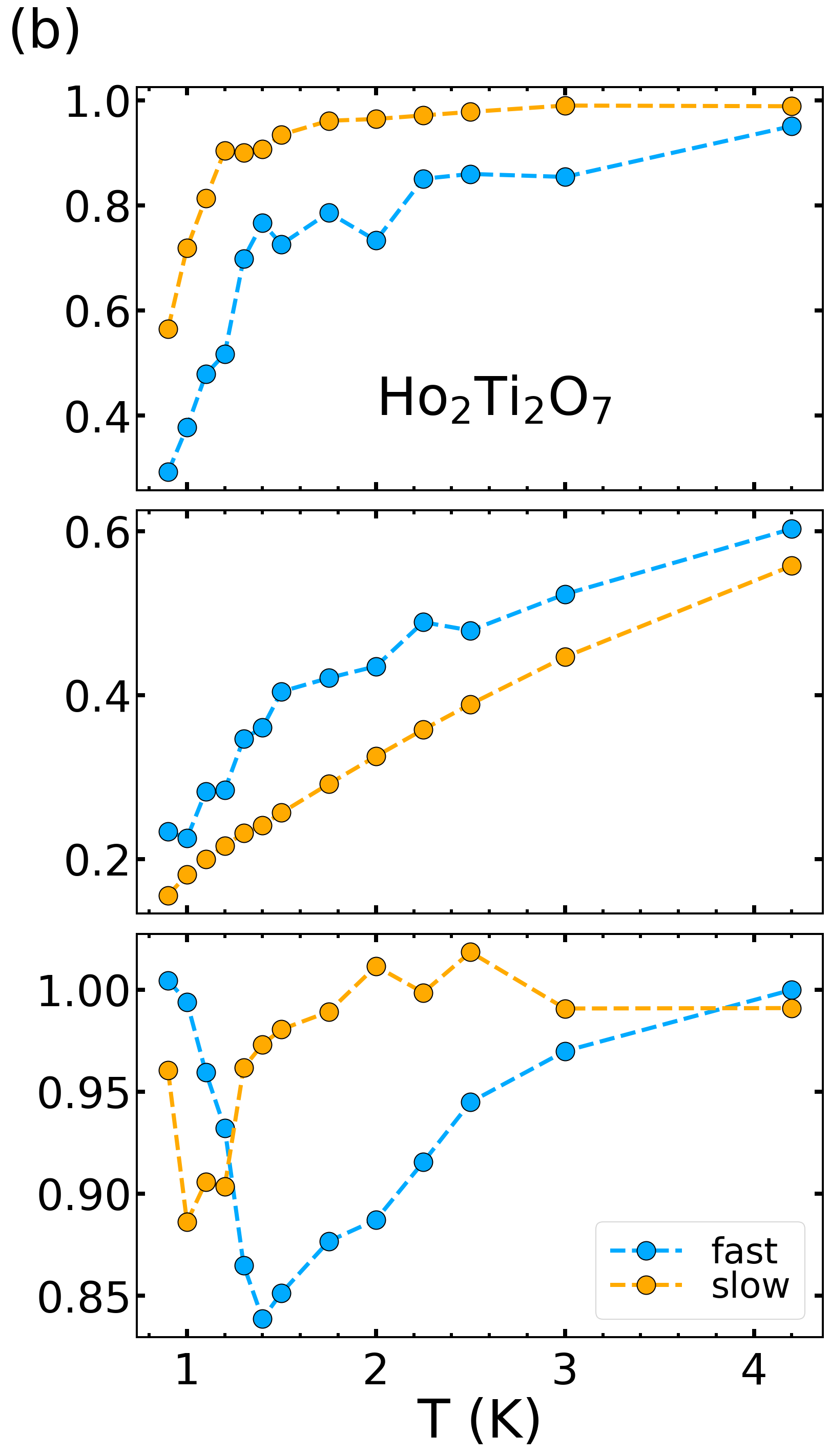}
\caption{\label{figparam} Relaxation time ratio $\tau_{\rm meas}/\tau_{\rm int}$ (top), noise at zero frequency ratio $S_{\rm 0,meas}/S_{\rm 0,int}$ (middle) and exponent ratio $\alpha_{\rm meas}/\alpha_{\rm int}$ (bottom) for (a) the single time in \dto\  with $N_{\rm DTO}=0.105\times4\pi$ and (b) the fast and slow processes in \hto\ with $N_{\rm s}=1/3\times4\pi$.
}
\end{figure}

Figure~\ref{figparam} shows the temperature evolution of the fitted parameters for both samples. In both compounds, the demagnetization effect induces a modest decrease in the $\tau$ and $S_0$ ratios as the temperature decreases from 4 to 2~K. Upon further cooling, the ratios decrease more sharply. This temperature dependence reflects the direct link between the magnitude of the demagnetization effect and the real part of the ac susceptibility $\chi'$, which increases continuously with decreasing temperature.
As expected from the fact that $N_{\rm s}$ is more than twice $N_{\rm DTO}$, the demagnetization effect is stronger for \hto\ than for \dto. At 800~mK, $\tau_{\rm meas}$ is less than half of $\tau_{\rm int}$ for \dto, and less than one third for \hto. The discrepancy is even more pronounced for the $S_{\rm 0,meas}/S_{\rm 0,int}$ ratio, which reaches 0.15 for \hto\ compared to 0.3 for \dto.

The fitted exponent $\alpha$ is also clearly affected by demagnetization corrections in both \dto\ and \hto, but the effect appears to be more complex as illustrated in the bottom panels of Figure \ref{figparam}. 
Below 1 K, the ratio $\alpha_{\rm meas}/\alpha_{\rm int}$ is close to 1. The characteristic times being very large at these low temperatures, the fitted frequency range is large enough to determine $\alpha$ accurately. 
When the temperature increases, this ratio sharply decreases down to 0.9 in \dto\ and 0.85 in \hto. This arises because the finite frequency range of our measurements limits access to the asymptotic high-frequency slope when the characteristic relaxation time decreases. As a result, $\alpha_{\rm meas}$ is fitted in an intermediate frequency region where the low-frequency regime influences the apparent slope.  After applying the demagnetization correction to the noise spectra, the fitted $\alpha_{\rm int}$ increases, as both the plateau value $S_0$ and the characteristic time $\tau$ are enhanced.
Nevertheless, the ratio starts increasing for temperatures above 1.5 K. Indeed, together with the $\tau$ diminution, the susceptibility decreases and so do the demagnetization corrections, reducing the difference between $\alpha_{\rm meas}$ and $\alpha_{\rm int}$. 
In the case of \hto, the effect on $\alpha_{\rm slow}$ is weaker than on $\alpha_{\rm fast}$, likely because in this frequency regime the fit is better constrained and less susceptible to the experimental artefacts described above.  

Notably, for \dto,  fits of the intrinsic noise yield $\alpha$ values closer to those reported by Billington et al.~\cite{Billington25}, obtained from ac susceptibility measurements in the 2--30~K range. This resolves the discrepancy between the $\alpha$ values extracted from ac susceptibility - which are close to theoretical expectations - and the anomalously low values reported in noise measurements \cite{Dusad19, Samarakoon22}. The latter were not corrected for demagnetization effects, which were not considered at the time to influence magnetic noise measurements, whereas the measurement geometries in those studies appear to favor significant demagnetization. As shown above, fitting the uncorrected noise in such geometries tends to underestimate $\alpha$ relative to the intrinsic value. Consequently, when fitting Monte Carlo simulations to experimental data that have not been corrected for demagnetizing effects, the measurement geometry must be explicitly incorporated to obtain a reliable parametrization of the spin ice Hamiltonian.

Importantly, despite these sizeable quantitative corrections, the overall temperature dependence of the dynamical parameters remains qualitatively unchanged. 
This is because $\tau$ and $S_0$ span several orders of magnitude over the measurement temperature range (see Appendix~\ref{fit_param}), so that the qualitative picture of the spin ice dynamics is preserved even when demagnetization effects are not accounted for.
Nevertheless, the demagnetization correction is most critical for the accurate determination of $\alpha$, which is susceptible to systematic underestimation. This is particularly significant given that $\alpha$ is a key parameter for understanding how correlations and constraints develop in the material, and is directly connected to the nature of the monopole dynamics~\cite{Hallen22}.

\section{Conclusion} \label{Conclusion}
In this work we have presented measurements of spectral noise density and ac susceptibility performed in a custom-built SQUID magnetometer equipped with a dilution refrigerator, spanning the temperature range 4.2~K to 600~mK. Experiments were carried out on two canonical spin ice materials, \hto\ and \dto, across five sample configurations of varying geometry. We establish the role of the demagnetizing factor $N$ in shaping both the measured ac susceptibility and the noise spectra.
We demonstrate that, for systems exhibiting a broad dynamical frequency response, such as spin ice, the intrinsic noise spectra cannot be straightforwardly inferred from noise measurements alone. Nonetheless, by exploiting the fluctuation-dissipation theorem, which directly relates the noise PSD to the ac susceptibility for which demagnetization corrections are tractable we have quantitatively assessed the impact of the demagnetizing field on the measured noise spectra. Using empirical fitting with phenomenological models, we have evaluated the effect of demagnetization on the dynamical parameters extracted from spin ice noise measurements. We find that while the qualitative behavior of the spin ice dynamics is preserved, the quantitative correction factors are substantial, reaching up to one order of magnitude for $S_0$ and a factor of three for $\tau$. 

The implications of this work extend beyond spin ice, as magnetic noise spectroscopy is becoming an increasingly widely used probe of the dynamics of frustrated and correlated magnetic systems, including spin liquids. A key motivation for noise measurements is the possibility of directly comparing the measured PSD with Monte Carlo simulations~\cite{Samarakoon22, Hallen22}. Our results demonstrate that such quantitative comparisons are more subtle than previously appreciated. Accurate benchmarking against theoretical models requires either extracting the intrinsic noise spectrum from experiment, performing simulations that account for the sample geometry, or designing experiments that minimize demagnetization effects through an appropriate choice of sample shape. Considering a broader perspective, our work suggests that demagnetizing effects should also be expected to play a role in other types of magnetization noise measurements, such as SQUID microscopy~\cite{Watson19}, NV-center magnetometry~\cite{Rovny22}, and Kerr magnetometry~\cite{Goryca21}. In these techniques, the primary objective is often to probe local magnetic objects and textures, for which demagnetizing corrections are not necessarily relevant. However, care should be taken when using these techniques to investigate bulk magnetic properties, as demagnetizing effects then become significant.

\bigskip
\acknowledgements
F. Morineau acknowledges financial support from the LANEF PhD Program.
We gratefully acknowledge G. Garde, A. G\'erardin, G. Pont and O. Tissot for their technical support in the development of the noise magnetometer.
We are very grateful to S. Bramwell and L. Bovo for providing us their spherical \hto\ sample. We also thank P. Holdsworth and S. Bramwell for fruitful discussions. 
The work at the University of Warwick and Cardiff University was supported by EPSRC, UK, through Grant EP/T005963/1 and EP/S016465/1 respectively.

\appendix
\section{Kramers-Kronig relation to correct noise for demagnetization}
\label{Kramers-Kronig}

When measuring the magnetic noise $S$ in a system at thermal equilibrium, the FDR provides direct access to the dissipative part of the susceptibility $\chi''$. However, correcting for demagnetizing effects requires knowledge of both $\chi'_{\rm meas}$ and $\chi''_{\rm meas}$, since the intrinsic imaginary susceptibility $\chi''_{\rm int}$ depends on both components through Equation~\ref{chi2_demag}.
In physical systems, the real and imaginary parts of any causal linear response function  such as the complex magnetic susceptibility $\chi$  are related through the Kramers--Kronig (KK) relations:

\begin{equation}
\begin{cases}
\chi'(\omega) =&\dfrac{1}{\pi} \, \mathcal{P} \displaystyle \int_{-\infty}^{\infty} \dfrac{\chi''(\omega')}{\omega' - \omega} \, d\omega' \\ \\
\chi''(\omega) =&-\dfrac{1}{\pi} \, \mathcal{P} \displaystyle \int_{-\infty}^{\infty} \dfrac{\chi'(\omega')}{\omega' - \omega} \, d\omega'
\end{cases}
\end{equation}
where $\mathcal{P}$ is the Cauchy principal value. 

In principle, the KK relations can be used to compute $\chi_{\rm noise}'(f)$ from the noise spectrum $S(f)$ by first converting $S$ into $\chi_{\rm noise}''$ via the FDR, and then applying the KK integration to obtain $\chi_{\rm noise}'$. Demagnetization corrections could then be performed on $\chi_{\rm noise}$ in the usual way, and the intrinsic noise recovered by applying the FDR to the corrected $\chi_{\rm noise}''$. In practice, however, this approach requires knowledge of $\chi''$ over the complete frequency range where it is non-zero, a condition that is difficult to satisfy experimentally, as illustrated below.

Figure~\ref{KK_demag_noise_0.8K} (left panels) shows $\chi''_{\rm noise}$ extracted from noise measurements at $T = 800$~mK via the FDR, together with $\chi'_{\rm noise}$ obtained by numerical KK integration of $\chi''_{\rm noise}$ (green lines), performed using the {\it pykk} Python library. These are compared with the directly measured ac susceptibility $\chi_{\rm meas}$ (green dots). As expected from the FDR, $\chi''_{\rm noise}$ and $\chi''_{\rm meas}$ are in excellent agreement. However, the KK-integrated $\chi'_{\rm noise}$ is slightly underestimated relative to the directly measured $\chi'_{\rm meas}$. This discrepancy arises from the limited frequency range of the measurements, which does not fully capture the spectral weight of $\chi''_{\rm noise}$. At this temperature, the missing spectral weight lies at low frequencies.

\begin{figure}[t!]
\begin{center}
\includegraphics[width=8.8cm]{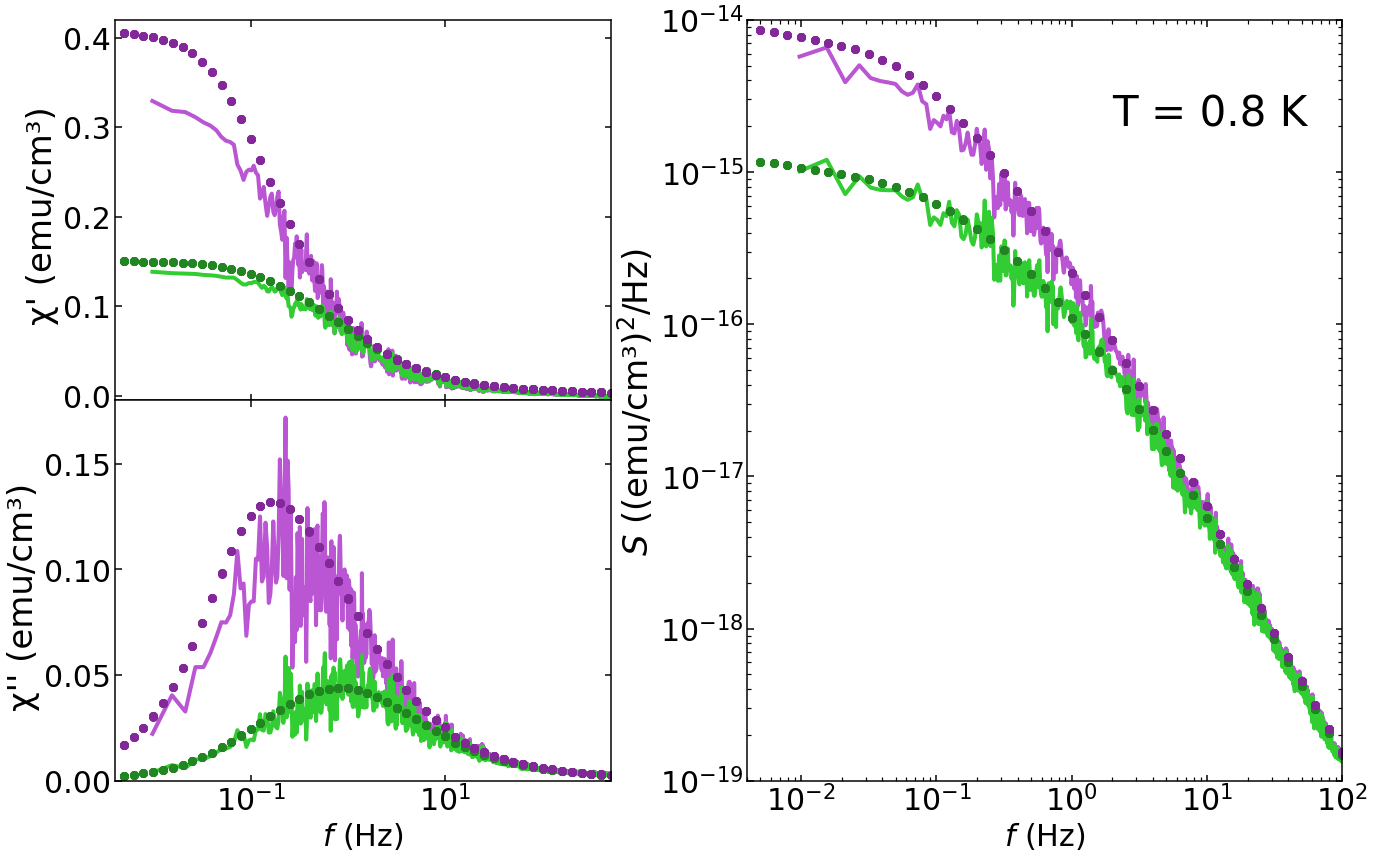}
\end{center}
\caption{\label{KK_demag_noise_0.8K} Demagnetization corrections on the sphere sample of \hto\ at 800 mK. Left panel: Measured (green) and intrinsic (purple) real part $\chi'$ (top) and imaginary part $\chi''$ (bottom) of the susceptibility vs frequency. Dots correspond to the data obtained from the ac susceptibility. Lines correspond to the ones obtained from the measured noise using the FDR and the KK relation (see text). Intrinsic values are obtained by performing demagnetization corrections (Equation \ref{chi2_demag}).
Right panel: Measured (green) and intrinsic (purple) noise PSD $S$ (lines) and the associated $D$ obtained from susceptibility (dots) vs frequency. }
\end{figure}

\begin{figure}[b]
\begin{center}
\includegraphics[width=8.8cm]{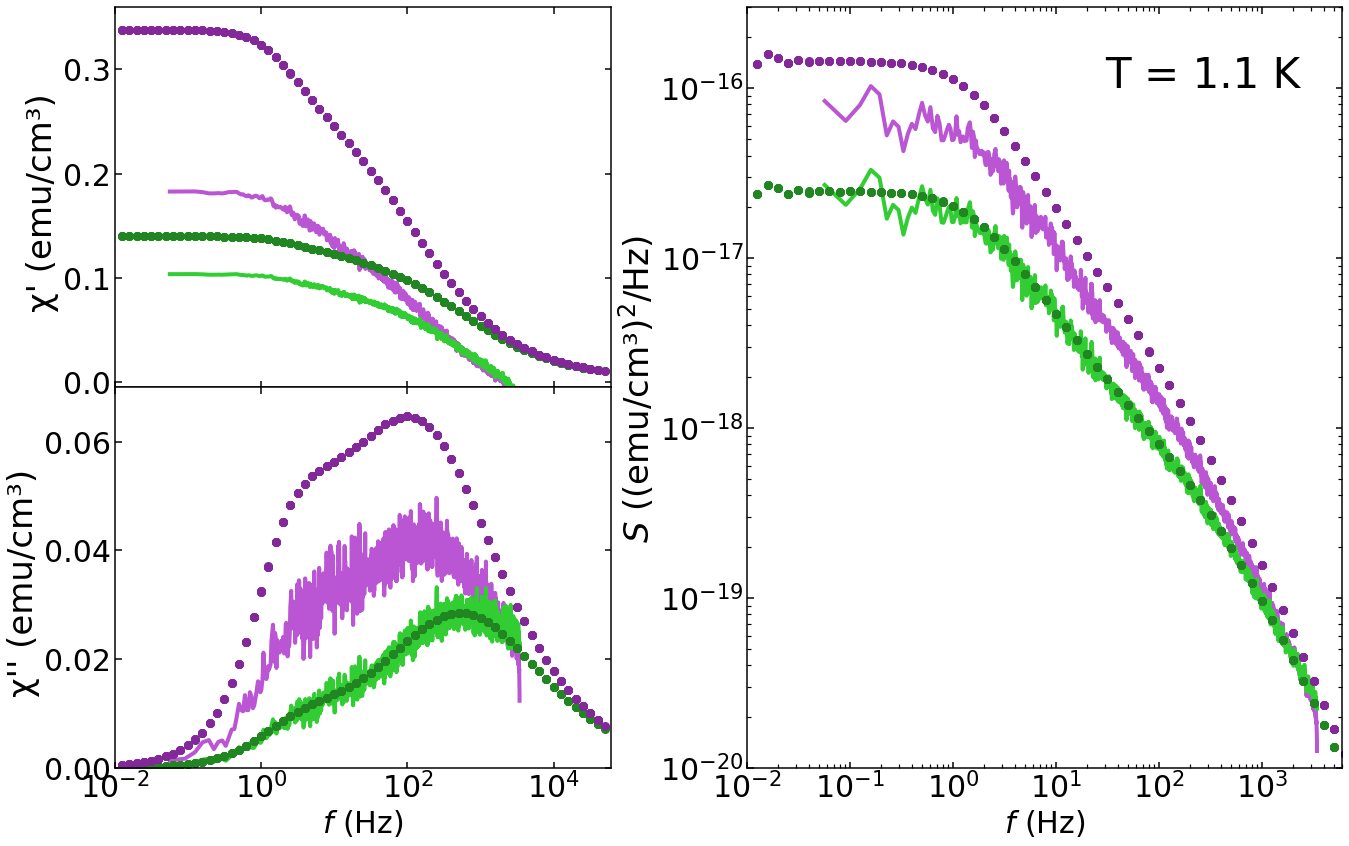}
\end{center}
\caption{\label{KK_demag_noise_1.1K} Demagnetization corrections on the sphere sample of \hto\ at 1.1 K. Left panel: Measured (green) and intrinsic (purple) real part $\chi'$ (top) and imaginary part $\chi''$ (bottom) of the susceptibility vs frequency. Dots correspond to the data obtained from the ac susceptibility. Lines correspond to the ones obtained from the measured noise using the FDR and the KK relation (see text). Intrinsic values are obtained by performing demagnetization corrections (Equation \ref{chi2_demag}).
Right panel: Measured (green) and intrinsic (purple) noise PSD $S$ (lines) and the associated $D$ obtained from susceptibility (dots) vs frequency.}
\end{figure}

\label{fit_param}
\begin{figure}[t!]
\begin{center}
\includegraphics[width=8cm]{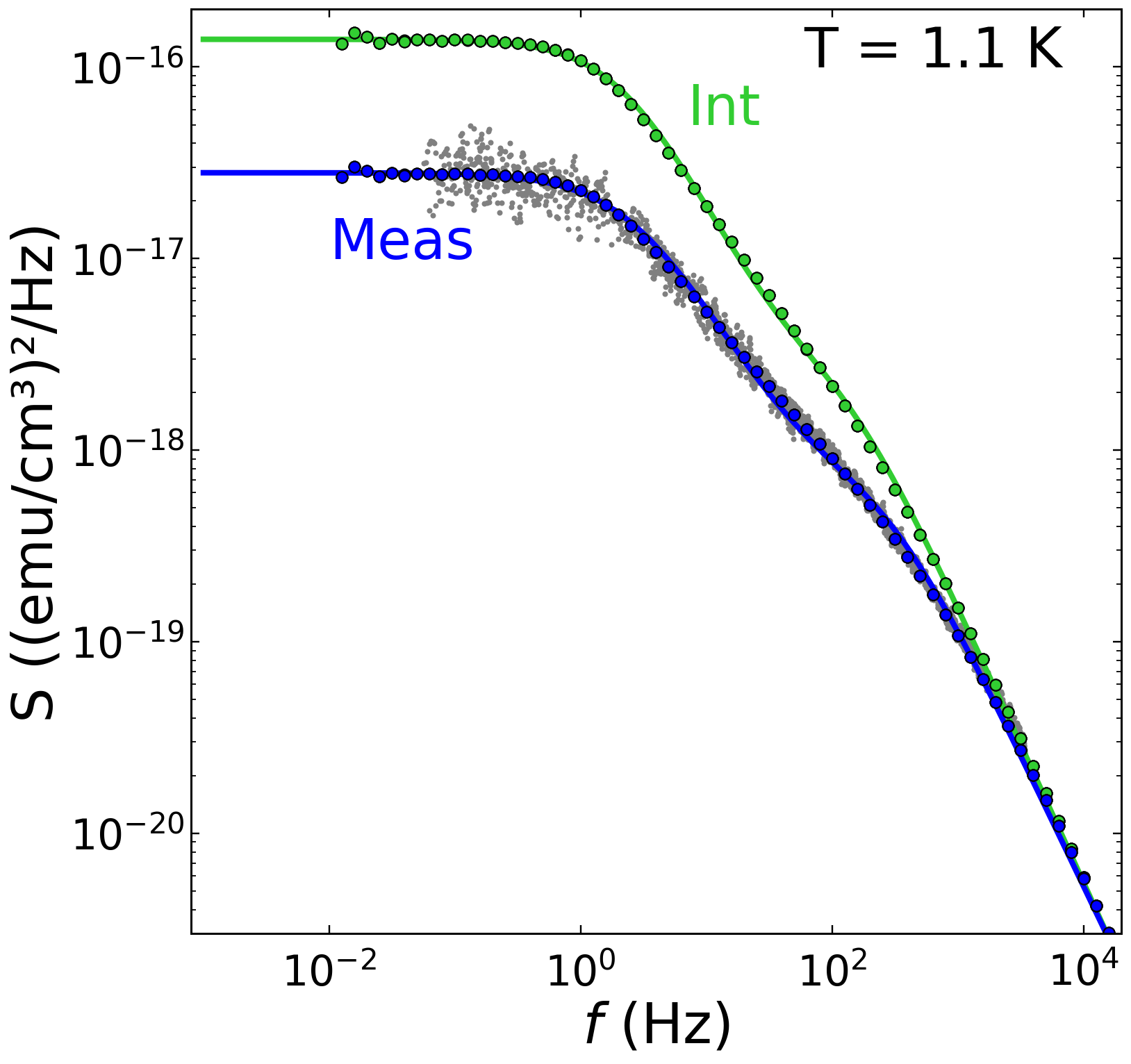}
\end{center}
\caption{\label{sphere_demag_1.1} Measured (blue dots) and intrinsic (green dots) $D(f)$ together with the measured noise spectra $S_{\rm meas}(f)$ (small grey dots) for the \hto\ sphere sample at 1.1 K. The blue and green solid lines indicate fits of $S_{\rm meas}(f)$ and $D_{\rm int}(f)$ respectively using Equation~\ref{Davidson_Cole_noise}. }
\end{figure}

\begin{figure}[b!]
\begin{center}
\includegraphics[height=6cm]{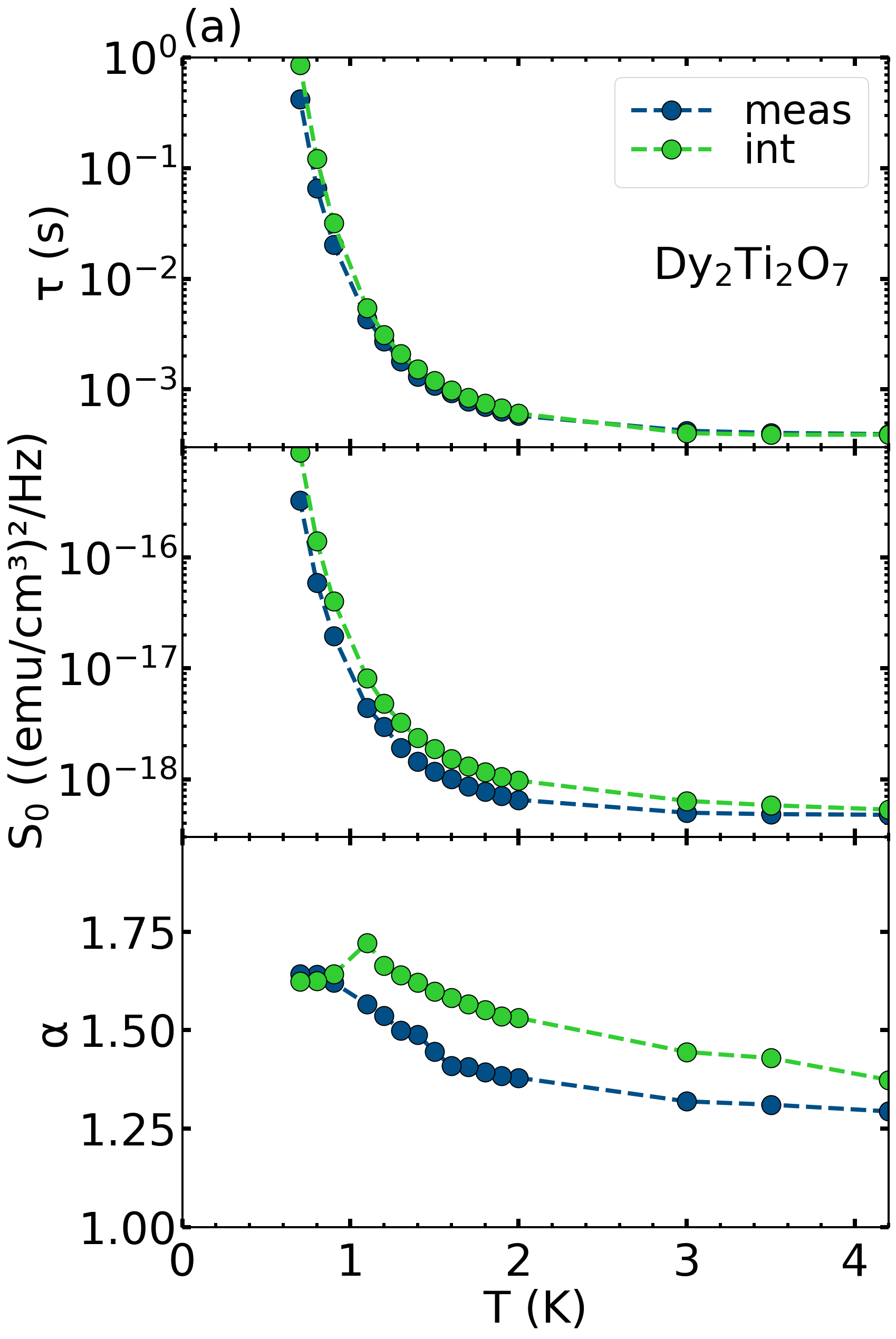}
\includegraphics[height=6cm]{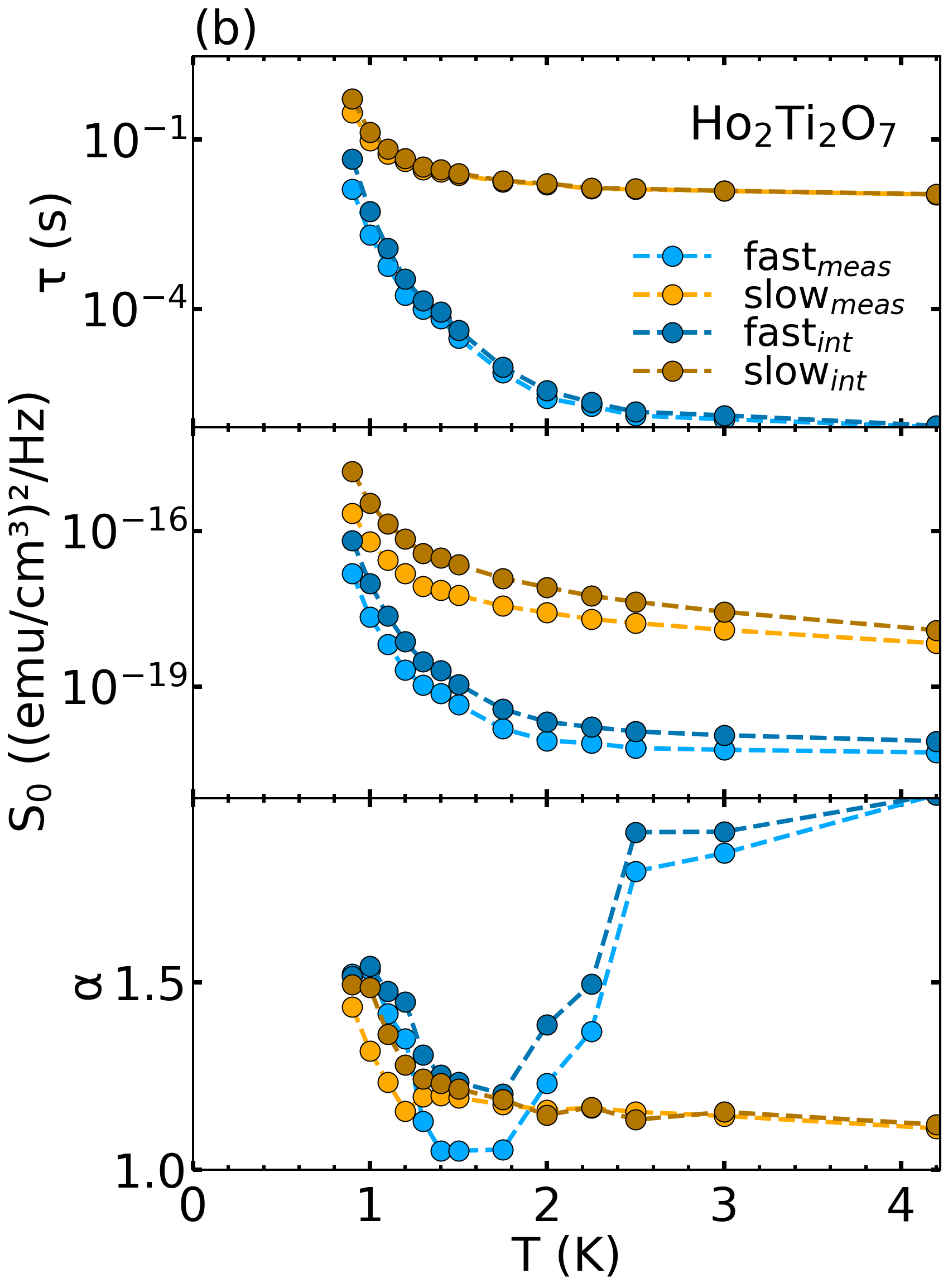}
\end{center}
\caption{\label{dto_param}Relaxation times $\tau_{\rm int}$ and $\tau_{\rm meas}$ (top), noise at zero frequency $S_{\rm 0,int}$ and $S_{\rm 0,meas}$ (middle) and $\alpha_{\rm int}$ and $\alpha_{\rm meas}$ parameters (bottom) obtained from the fits of $D_{\rm int}(f)$ and $S_{\rm meas}(f)$ respectively with (a) Equation~\ref{Davidson_Cole_noise} for the \dto\ sample with $N_{\rm DTO}=0.105\times4\pi$ and with (b) a sum of two contributions of the type of Equation~\ref{Davidson_Cole_noise} for the \hto\ sphere sample with $N_{\rm s}=1/3 \times4\pi$.}
\end{figure}

When the demagnetization correction is applied to $\chi_{\rm noise}$, constructed from the KK-derived $\chi'_{\rm noise}$ and the FDR-derived $\chi''_{\rm noise}$, both $\chi'_{\rm int}$ and $\chi''_{\rm int}$ (purple lines) are underestimated relative to the intrinsic susceptibility obtained directly from ac susceptibility measurements (purple dots). This is expected, as the spectral weight loss in $\chi'_{\rm noise}$ is amplified by the demagnetization correction. Nevertheless, because the loss remains small at this temperature, Figure~\ref{KK_demag_noise_0.8K} (right panel) shows that the resulting intrinsic noise spectrum $S_{\rm int}(f)$ remains close to the expected curve at 800~mK.

This KK correction protocol performs reasonably well when the full relaxation process falls within the experimental frequency window. Its limitations become apparent, however, as soon as the relaxation process shifts outside the accessible frequency range, as occurs at other measured temperatures. This is illustrated in Figure~\ref{KK_demag_noise_1.1K}, which shows the same quantities as Figure~\ref{KK_demag_noise_0.8K} but at $T = 1.1$~K. At this temperature, the relaxation response shifts to higher frequencies and the noise reaches the SQUID noise floor at approximately $f = 3.2$~kHz. Beyond this frequency, the signal is no longer accessible, and since these high frequencies carry a substantial fraction of the spectral weight, the KK integration suffers a more pronounced loss of intensity in $\chi'_{\rm noise}$. This loss propagates through the demagnetization correction, leading to a more severe underestimation of both $\chi'_{\rm int}$ and $\chi''_{\rm int}$. As shown in the right panel of Figure~\ref{KK_demag_noise_1.1K}, the corrected noise obtained by this method deviates substantially from the intrinsic noise and is underestimated by a factor of two at low frequencies.

Since the relaxation times in spin ice evolve over several decades with temperature~\cite{Snyder04, Matsuhira11, Quilliam11, Wang21}, achieving full spectral coverage across the entire temperature range is generally not possible. While the KK-based correction method is sound in principle, it requires complete spectral coverage that magnetic noise and ac susceptibility experiments typically cannot provide.

\FloatBarrier
\section{Fit parameters}
The procedure used to extract the intrinsic and measured fit parameters is illustrated in Figure~\ref{sphere_demag_1.1} for the \hto\ sphere sample at 1.1~K. The intrinsic noise spectrum $D_{\rm int}(f)$ (green dots) is obtained from the ac susceptibility corrected for demagnetization effects and converted to a noise spectrum via the FDR. Both $D_{\rm int}(f)$ and the measured noise spectrum $S_{\rm meas}(f)$ (grey dots) are fitted to a sum of two contributions of the form of Equation~\ref{Davidson_Cole_noise}. For \dto, a single contribution is used.

Figures~\ref{dto_param}~(a) and (b) show the temperature dependence of the $\tau$, $S_0$ and $\alpha$ parameters obtained by this procedure for \dto\ and \hto, respectively. These parameters are used to compute the temperature dependence of the $\tau$, $S_0$ and $\alpha$ ratios shown in Figure~\ref{figparam} of the main text.

For \hto, note that the maximum errors resulting from the uncertainty on the sphere ac susceptiblity are of 10 \% for $S_0$, 5 \% for $\tau_0$, 1 \% for $\alpha_0$, 17 \% for $S_1$, 3 \% for $\tau_1$ and 1.5 \% for $\alpha_1$.


\bibliography{biblio}

\end{document}